\documentclass[
 reprint,twocolumn,
 amsmath,amssymb,
 aps,
]{revtex4}

\usepackage{epsfig,bm,epsf,graphics}
\usepackage{graphicx}% Include figure files
\usepackage{dcolumn}% Align table columns on decimal point
\usepackage{bm}% bold math
\begin{document}
%\draft
\preprint{APS/123-QED}

\title{Partial phase transition and quantum effects in helimagnetic films under an applied magnetic field}% Force line breaks with \\

\author{Sahbi El Hog\footnote{sahbi.el-hog@u-cergy.fr} and H. T. Diep\footnote{diep@u-cergy.fr, corresponding author}}
\affiliation{%
Laboratoire de Physique Th\'eorique et Mod\'elisation,
Universit\'e de Cergy-Pontoise, CNRS, UMR 8089\\
2, Avenue Adolphe Chauvin, 95302 Cergy-Pontoise Cedex, France.\\
 }%


\begin{abstract}
We study the phase transition in a helimagnetic film with Heisenberg spins under an applied magnetic field in the $c$ direction perpendicular to the film.
The helical structure is due to the antiferromagnetic interaction between next-nearest neighbors in the $c$ direction. Helimagnetic films in zero field are known to have a strong modification of the in-plane helical angle near the film surfaces. We show that spins react to a moderate applied magnetic field by creating a particular spin configuration along the $c$ axis.  With increasing temperature ($T$), using Monte Carlo simulations we show that the system undergoes a phase transition triggered by the destruction of the ordering of a number of layers. This partial phase transition is shown to be intimately related to the ground-state spin structure. We show why some layers undergo a phase transition while others do not. The Green's function method for non collinear magnets is also carried out to investigate effects of quantum fluctuations.  Non-uniform zero-point spin contractions and a crossover of layer magnetizations  at low $T$ are shown and discussed.
\vspace{0.5cm}
\begin{description}
\item[PACS numbers: 75.25.-j ; 75.30.Ds ; 75.70.-i ]
\end{description}
\end{abstract}

\pacs{Valid PACS appear here}% PACS, the Physics and Astronomy
\maketitle

%\tableofcontents

\section{Introduction}
\begin{figure}[ht!]
\centering
\includegraphics[width=5cm,angle=0]{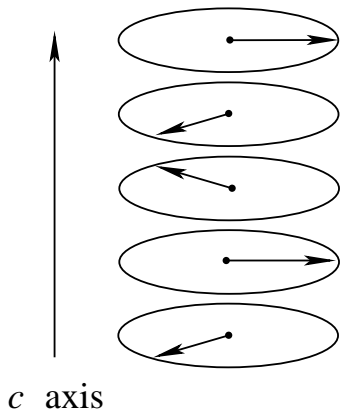}
\caption{Spin configuration along the $c$ direction in the bulk case where $J_2/J_1=-1$, $H/J_1=0$. \label{helical}}
\end{figure}

Helimagnets have been subject of intensive investigations over the last four decades since the discovery of its ordering \cite{Yoshimori,Villain59}: in the bulk, a spin in a space direction turns an angle $\theta$ with respect to the orientation of its previous nearest neighbor (see Fig. \ref{helical}). This helical structure can take place in several directions simultaneously with different helical angles. The helical structure shown in Fig. \ref{helical} is due to the competition between the interaction between nearest neighbors (NN) and the antiferromagnetic interaction between next-nearest neighbors (NNN).
Other helimagnetic structures have also been very early investigated \cite{Bertaut1,Bertaut2,Bertaut3}.
Spin-wave properties in bulk helimagnets have been investigated by spin-wave theories \cite{Harada,Rastelli,Diep89} and Green's function method \cite{Quartu1998}. Heat capacity in bulk MnSi has been experimentally investigated \cite{Stishov}.

We confine ourselves to the case of a Heisenberg helical film in an applied magnetic field. Helimagnets are special cases of a large family of periodic non collinear spin structures called frustrated systems of XY and Heisenberg spins.
The frustration has several origins: (i) it can be due to the geometry of the lattice such as the triangular lattice, the face-centered cubic (FCC) and hexagonal-close-packed (HCP) lattices, with antiferromagnetic NN interaction
\cite{Bak,Plumer,Maleyev} (ii) it can be due to competing interactions between NN and NNN such as the case of helimagnets \cite{Yoshimori,Villain59} shown in Fig. \ref{helical}  (iii) it can be due to the competition between the exchange interaction which favors collinear spin configurations and the Dzyaloshinskii-Moriya  (DM) interaction which favors perpendicular spin arrangements.

%The family of helimagnets due to the DM interaction has attracted a growing interest of researchers:
%this interaction without inversion symmetry has been shown to generate skyrmions in some magnetic compounds which
%can have useful  applications in spin transport properties in thin films and multilayers \cite{Fert2013}.
%Intensive researches have been carried out to understand theoretically the origin and the role of
%skyrmions \cite{Lin,Bogdanov,Rossler,Muhlbauer,Yu1,Yu2,Seki,Adams}.
%Experimentally, there is a large number of investigations which has recently been performed on thin
%films of helical magnets \cite{Heurich,Wessely,Jonietz}.
%It should be emphasized that skyrmions can also be generated without the DM interaction as it
%has been recently shown in Refs. \onlinecite{Hayami2016,Okubo}. Thus, the existence of skyrmions
%in a system is not restricted to the DM interaction. Rather, it is due to the competition
%between various interactions each of which favors a different spin configuration.

Effects of the frustration have been extensively studied in various systems during the last 30 years. The reader is referred to recent reviews on bulk frustrated systems given in Ref. \onlinecite{DiepFSS}.
When frustration effects are coupled with surface effects, the situation is often complicated.  Let us mention our previous works on a frustrated surface \cite{NgoSurface} and on a frustrated FCC antiferromagnetic film \cite{NgoSurface2} where surface spin rearrangements and surface phase transitions have been found.
We have also recently shown results in zero field of thin films of body-centered cubic (BCC) and simple cubic (SC) structures \cite{Diep2015,Sahbi}.  The helical angle along the $c$ axis perpendicular to the film surface was found to strongly vary in the vicinity of the surface. The phase transition and quantum fluctuations have been presented.

In this paper we are interested in the effect of an external magnetic field applied along the $c$ axis perpendicular to the film surface of a helimagnet with both classical and quantum Heisenberg spins. Note that without an applied field, the spins lie in the $xy$ planes: spins in the same plane are parallel while two NN in the adjacent planes form an angle $\alpha$ which varies with the position of the planes  \cite{Sahbi}, unlike in the bulk.  As will be seen below, the applied magnetic field gives a very complex spin configuration  across the film thickness. We determine this ground state (GS) by the numerical steepest descent method.  We will show by Monte Carlo (MC) simulation that the phase transition in the field is due to the disordering of a number of layers inside the film.  We identify the condition under which a layer becomes disordered. This partial phase transition is not usual in thin films where one observes more often the disordering of the surface layer, not an interior layer. At low temperatures, we investigate effects of quantum fluctuations using a Green's function (GF) method for non-collinear spin configurations.

The paper is organized as follows. Section II is devoted to the description of the model and the determination of the classical GS. The structure of the GS spin configuration  is shown as a function of the applied field. Section III is used to show the MC results at finite temperatures where a partial phase transition is observed. Effects of the magnetic field strength and the film thickness are shown.
The GF method is described in section IV and its results on the layer magnetizations at low temperatures are displayed
and discussed in terms of quantum fluctuations. Section V is devoted to concluding remarks.

%Fig1

\section{Model - Determination of the classical ground state}\label{GSSC}

We consider a thin film of SC lattice of $N_z$ layers stacked in the $c$ direction.  Each lattice site is occupied by a Heisenberg spin. For the GS determination, the spins are supposed to be classical spins in this section. The Hamiltonian is given by

\begin{equation}
\mathcal H=-\sum_{\left<i,j\right>}J_{i,j}\mathbf S_i\cdot\mathbf S_j -\sum_i \mathbf H\cdot \mathbf S_i
 \label{eqn:hamil1}
\end{equation}
where $J_{i,j}$ is the interaction between two spins $\mathbf S_i$ and $\mathbf S_j$ occupying the lattice sites $i$ and $j$ and $\mathbf H$ denotes an external magnetic field applied along the $c$ axis.  To generate helical angles in the $c$ direction, we suppose an antiferromagnetic interaction $J_2$ between NNN in the $c$ direction in addition to the ferromagnetic interaction $J_1$ between NN in all directions. For simplicity, we suppose that $J_1$ is the same everywhere.
For this section we shall suppose $J_2$ is the same everywhere for the presentation clarity.  Note that in the bulk in zero field, the helical angle along the $c$ axis is given by $\cos \alpha=-\frac{J_1}{4J_2}$ for a SC lattice \cite{DiepTM}with $|J_2|>0.25 J_1$. Below this value, the ferromagnetic ordering is stable.

In this paper we will study physical properties as functions of $J_2/J_1$, $H/J_1$ and $k_BT/J_1$.  Hereafter, for notation simplicity we will take $J_1=1$ and $k_B=1$. The temperature is thus in unit of $J_1/k_B$, the field and the energy are in unit of $J_1$.

In a film, the angles between NN in adjacent planes are not uniform across the film: a strong variation is observed near the surfaces.  An exact determination can be done by energy minimization \cite{Diep2015} or by numerical steepest descent method \cite{NgoSurface,NgoSurface2}. The latter is particularly efficient for complex situations such as the present case where  the spins are no longer in the $xy$ planes in an applied field: a spin in the $i-th$ layer is determined by  two parameters which are the angle with its NN in the adjacent plane, say $\alpha_{i,i+1}$, and the azimuthal angle $\beta_i$ formed with the $c$ axis. Since there is no competing interaction in the $xy$ planes, spins in each plane are parallel. In this paper we use the steepest descent method which consists in calculating the local field at each site and aligning the spin in its local field to minimize its energy. The reader is referred to Ref. \onlinecite{NgoSurface} for a detailed description. In so doing for all sites and repeating many times until a convergence to the lowest energy is obtained with a desired precision (usually at the 6-th digit, namely at $\simeq 10^{-6}$ per cents), one obtains the GS configuration.
Note that we have used several thousands of different initial conditions to check the convergence to a single GS for each set of parameters.

Figures \ref{szsxy1}a and \ref{szsxy1}b show the spin components $S^z$, $S^y$ and $S^x$ for all layers.  The spin lengths in the $xy$ planes are shown in Fig \ref{szsxy1}c.  Since the spin structure in a field is complicated and plays an important role in the partial phase transition shown in the next section, let us describe it in details and explain the physical reason lying behind:

%Fig2

\begin{figure}[ht!]
\centering
\includegraphics[width=6cm,angle=0]{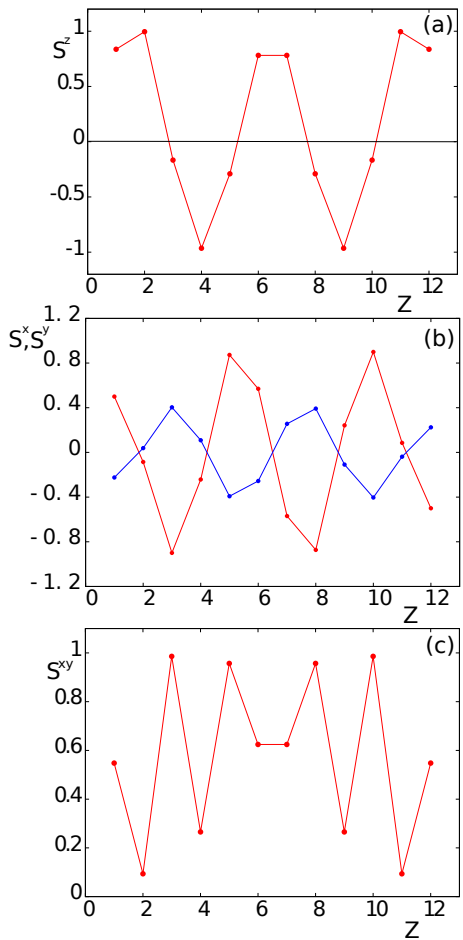}
\caption{ Spin components across the film in the case where $H=0.2$. The horizontal axis $Z$ represents plane $Z$ ($Z=1$ is the first plane etc.): (a) $S^{z}$ ; (b) $S^{x}$ (red) and $S^{y}$ (blue); (c) Modulus $S^{xy}$ of the projection of the spins on the $xy$ plane. See text for comments. \label{szsxy1}}
\end{figure}

\begin{itemize}

\item Several planes have negative $z$ spin components.
This can be understood by examining the competition between the magnetic field which tends to align spins in the $c$ direction, and the antiferromagnetic interaction $J_2$ which tries to preserve the antiferromagnetic ordering. This is very similar to the case of collinear antiferromagnets: in a weak magnetic field the spins remain antiparallel, and in a moderate field, the so-called "spin flop" occurs: the neighboring spins stay antiparallel with each other but turn themselves perpendicular to the field direction to reduce the field effect \cite{DiepTM}.

\item Due to the symmetry of the two surfaces,  one observes the following symmetry with respect to the middle of the film:

(i) $S_1^z=S_{N_z}^z$,  $S_2^z=S_{N_z-1}^z$, $S_3^z=S_{N_z-2}^z$ etc.

(ii) $S_1^y=-S_{N_z}^y$,  $S_2^y=-S_{N_z-1}^y$, $S_3^y=-S_{N_z-2}^y$ etc.

(iii) $S_1^x=-S_{N_z}^x$,  $S_2^x=-S_{N_z-1}^x$, $S_3^x=-S_{N_z-2}^x$ etc.

Note that while the $z$ components are equal, the $x$ and $y$ components are antiparallel (Fig. \ref{szsxy1}b): the spins preserve their antiferromagnetic interaction for the transverse components. This is similar to the case of spin flop in the bulk (see p. 86 of Ref. \onlinecite{DiepTM}).  Only at a very strong field that all spins turn into the field direction.

\item The GS spin configuration depends on the film thickness.  An example will be shown in the next section.

\end{itemize}

\begin{figure}[ht!]
\centering
\includegraphics[width=2.1cm,angle=0]{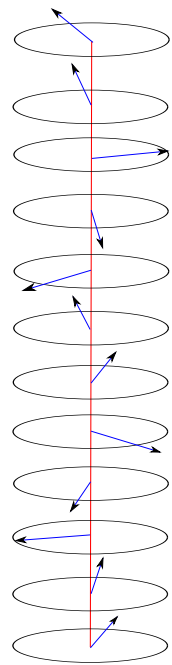}
\caption{ Spin configuration in the case where $H=0.4$, $J_2=-1$, $N_z=12$.  The circles in the $xy$ planes with radius equal to 1 are plotted to help identify the orientation of each spin.
The spins when viewed along the $c$ axis are shown in Fig. \ref{topview}d. \label{GSH01J2M1}}
\end{figure}

A full view of the "chain" of $N_z$ spins along the $c$ axis between the two surfaces is shown in Fig. \ref{GSH01J2M1}.

%Fig3

Note that the angle in the $xy$ plane is determined by the NNN interaction $J_2$. Without field, the symmetry is about the $c$ axis, so $x$ and $y$ spin components are equivalent (see Fig. 1). Under the field, due to the surface effect, the spins make different angles with the $c$ axis giving rise to different $z$ components for the layers across the film as shown in Fig. 2a. Of course, the symmetry axis is still the $c$ axis, so all $S^x$ and $S^y$ are invariant under a rotation around the $c$ axis. Figure 2b shows the symmetry of $S^x$ as that of $S^y$ across the film as outlined in remark (iii).  Fig. 2b is thus an instantaneous configuration between $S^x$ and $S^y$ for each layer across the film. As the simulation time is going on these components rotate about the $c$ axis but their symmetry outlined in remark (iii) is valid at any time.  The $xy$ spin modulus $S^{xy}$ shown in Fig. 2c, on the other hand, is time-invariant. The phase transition occurring for layers with large $S^{xy}$ ($xy$ disordering) is shown in the next section.

\begin{figure}[ht!]
\centering
\includegraphics[width=6cm,angle=0]{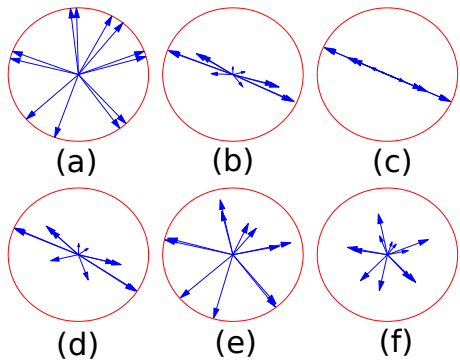}
\caption{ Top view of $S^{xy}$ (projection of spins on $xy$ plane) across the film for several values of
$H$: (a) 0, (b) 0.03, (c) 0.2, (d) 0.4, (e) 0.7, (f) 1.7. The radius of the circle, equal to 1, is the spin full length: for high fields, spins are strongly aligned along the $c$ axis, $S^{xy}$ is therefore much smaller than 1.\label{topview} }
\end{figure}

The GS spin configuration depends on the field magnitude $H$. If  $H$ increases, we observe an interesting phenomenon:  Figure \ref{topview} shows the spin configurations projected on the $xy$ plane (top view) for increasing magnetic field. We see that the spins of each chain tend progressively to lie in a same plane perpendicular to the $xy$ planes (Figs. \ref{topview}a-b-c).  The "planar zone" observed in Fig. \ref{topview}c occurs between $H\simeq 0.35$ and 0.5.  For stronger fields they are no more planar (Fig. \ref{topview}d-e-f).   Note that the larger the $xy$ component is, the smaller the $z$ component becomes: for example in Fig. \ref{topview}a the spins are in the $xy$ plane without field ($H=0$) and in Fig. \ref{topview}f they are almost parallel to the $c$ axis because of a high field.

%Fig4

\section{Phase transition}

We recall that for bulk materials,  in spite of their long history, the nature of the phase transition in non-collinear magnets such as stacked triangular XY and Heisenberg antiferromagnets has been elucidated only recently \cite{Diep89b,Ngo08,Ngo09}.  On the other hand, surface effects in thin films have
been intensively studied during the last three decades \cite{Heinrich,Zangwill,DiepTM}. Most of theoretical studies were limited to collinear magnetic orderings. Phase transitions in thin films with non-collinear ground states have been only recently studied \cite{NgoSurface,NgoSurface2,Diep2015,Sahbi,Mello2003}. MC simulations of a helimagnetic thin film \cite{Cinti2008} and a few experiments in helimagnets \cite{Karhu2011,Karhu2012} have also been carried out. These investigations were motivated by the fact that helical magnets present a great potential of applications in spintronics with spin-dependent electron transport \cite{Heurich,Wessely,Jonietz}.

As described in the previous section, the planar helical spin configuration in zero field becomes non planar in a perpendicular field.  In order to interpret the phase transition shown below, let us mention that a layer having a large $z$ spin-component parallel to the field cannot have a phase transition because its magnetization will never become zero. This is  similar to a ferromagnet in a field.
However, layers having large negative $z$ spin-components (antiparallel to the field) can undergo a transition due to the magnetization reversal at a higher temperature similarly to an antiferromagnet in a field.  In addition, the $xy$ spin-components whose $xy$ fluctuations are  not affected by the perpendicular field can make a transition.  Having mentioned these, we expect that some layers will undergo a phase transition, while others will not. This is indeed what we observed in MC simulations shown in the following.

For MC simulations, we use the Metropolis algorithm (see chapter 8 of Ref. \onlinecite{DiepTM}) and a sample size $N\times N\times N_z$ with $N=20$, 40, 60, 100 for detecting lateral-size effects and $N_z=8$, 12, 16 for thickness effects.  The equilibrium time is $10^5$ MC steps/spin and the thermal average is performed with the following $10^5$ MC steps/spin.

\subsection{Results: example of 12-layer film}

\begin{figure}[ht!]
\centering
\includegraphics[width=6cm,angle=0]{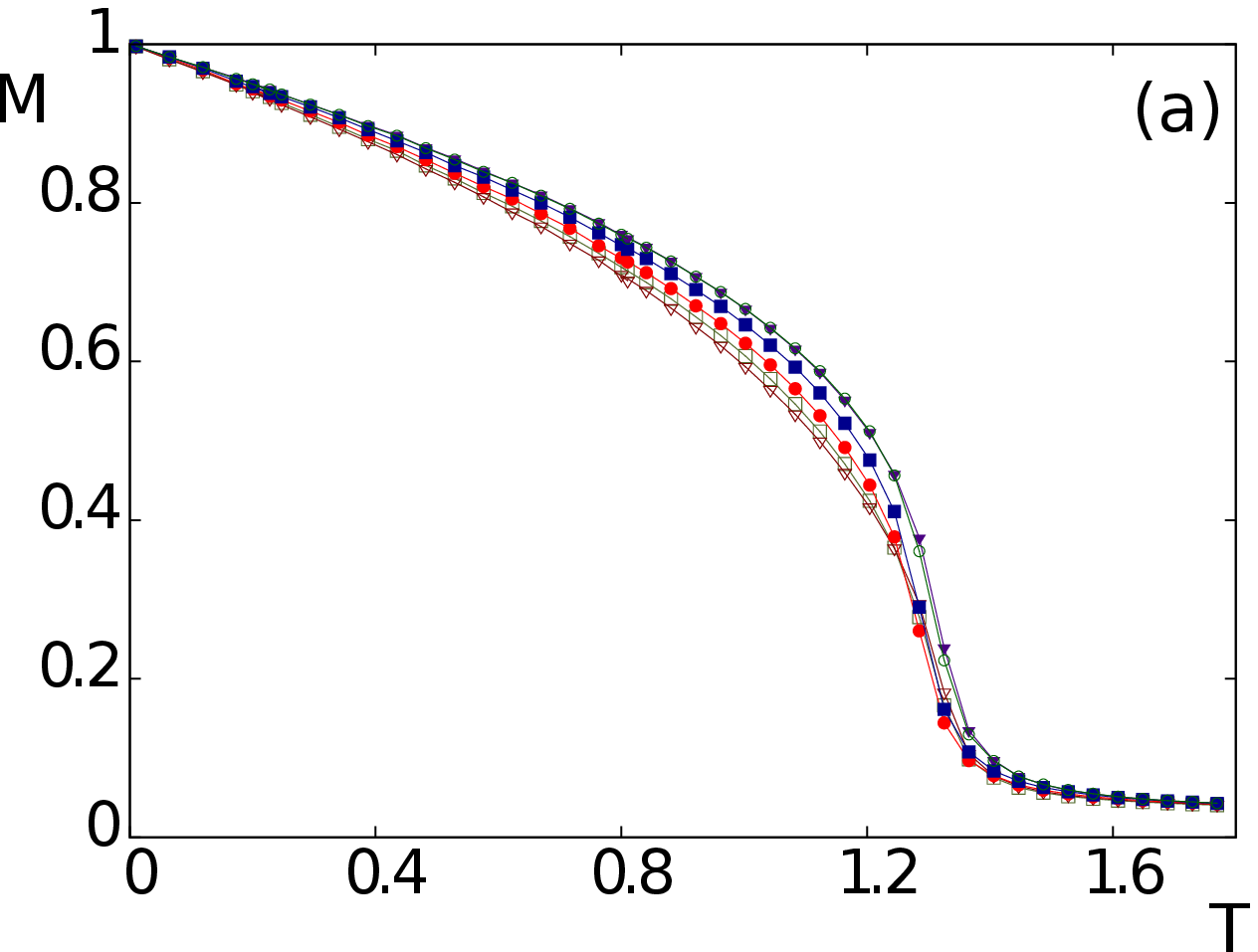}
\includegraphics[width=6cm,angle=0]{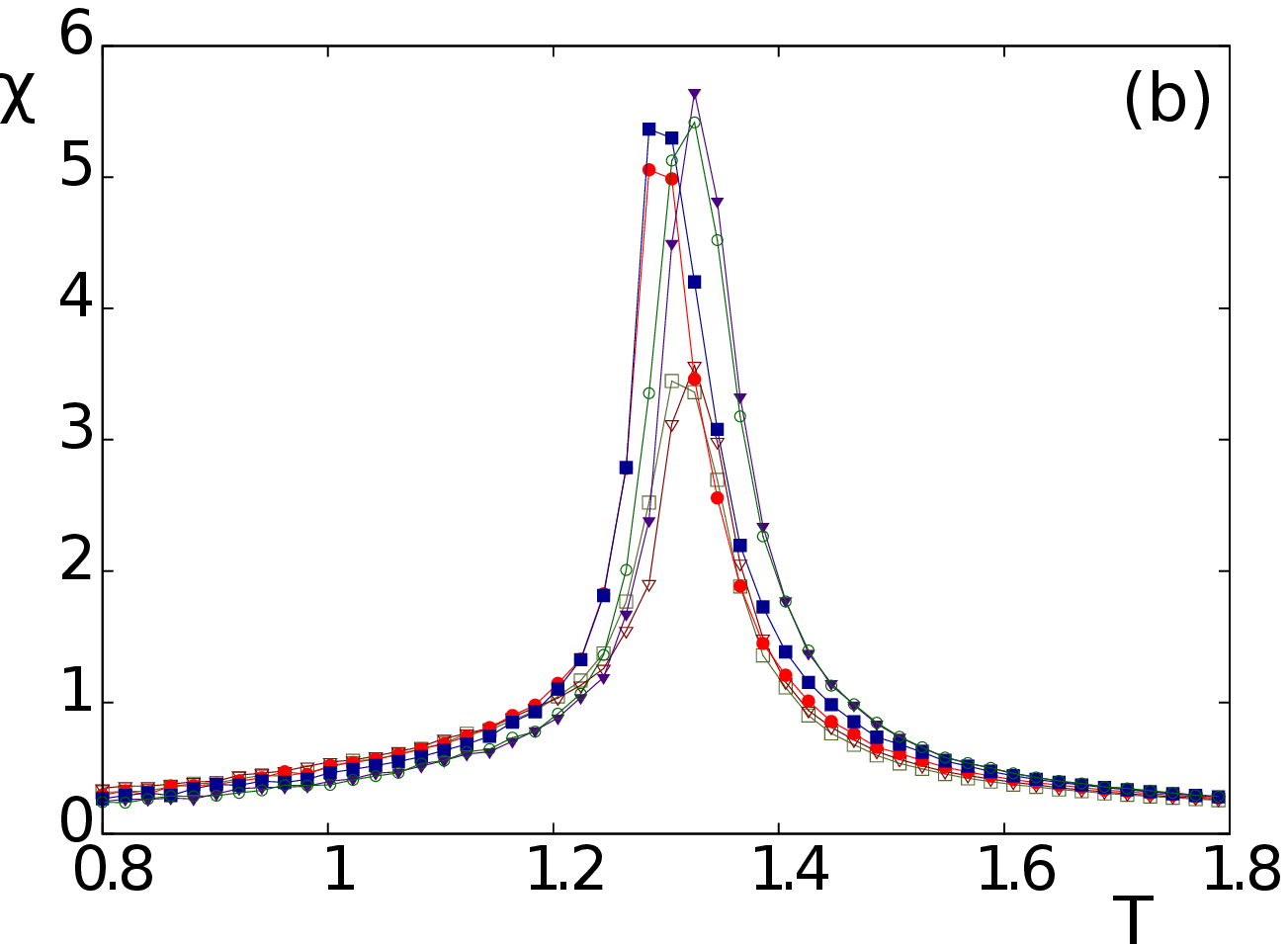}
\caption{(a) Layer magnetization and (b) layer magnetic susceptibility versus $T$
for  $H=0$, $J_2=-1$, $N_z=12$. Dark olive green void squares for the first layer, maroon void
triangles for the second, red circles for the third, indigo triangles for the fourth, dark blue squares for the fifth, dark green void circles for the sixth layer. \label{MXT1} }
\end{figure}

In  order to appreciate the effect of the applied field, let us show first the case where $H=0$ in Fig. \ref{MXT1}. We see there that all layers undergo a phase transition within a narrow region of $T$.
%Fig5

%Fig6
\begin{figure}[ht!]
\centering
\includegraphics[width=6cm,angle=0]{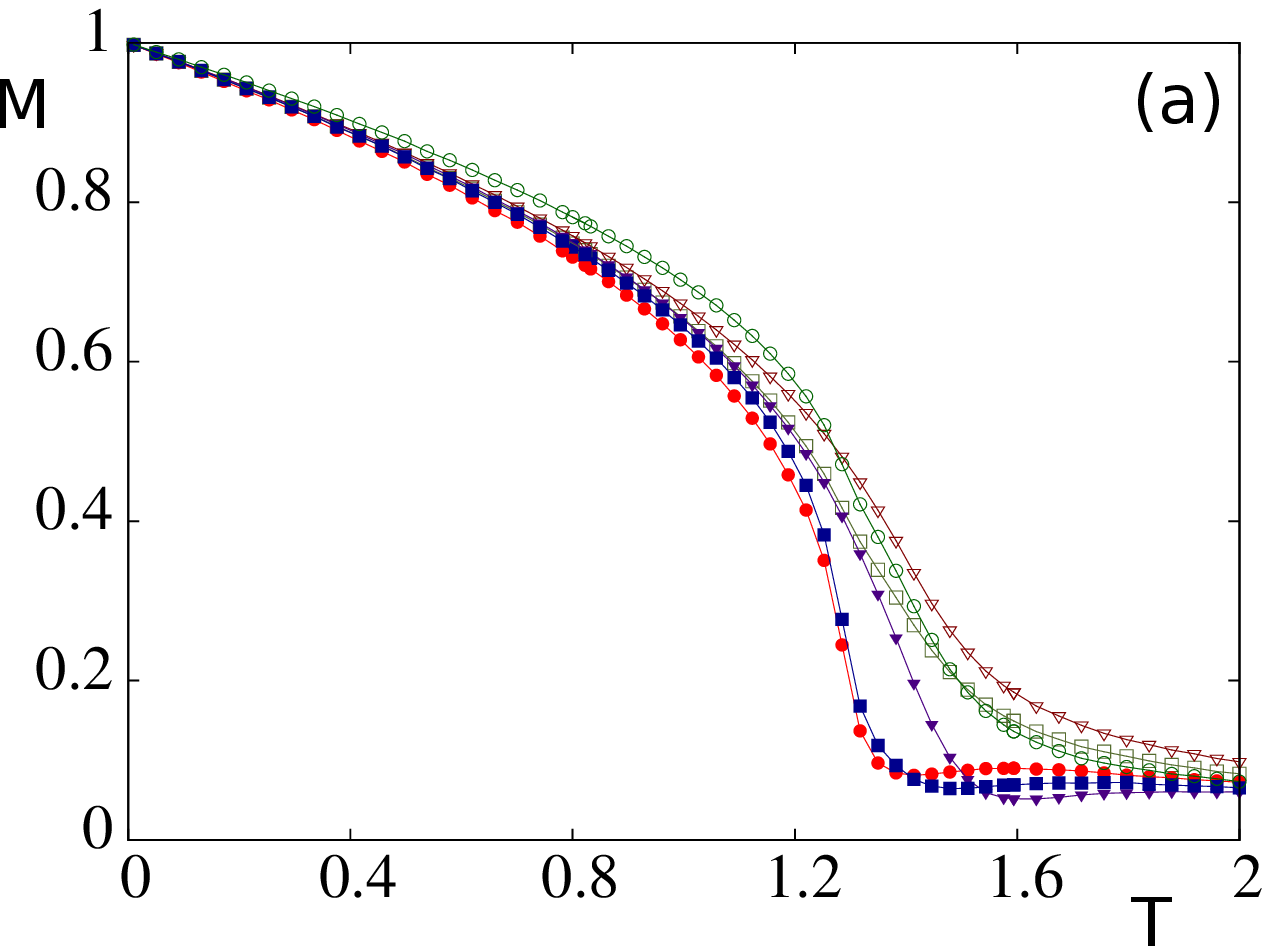}
\includegraphics[width=6cm,angle=0]{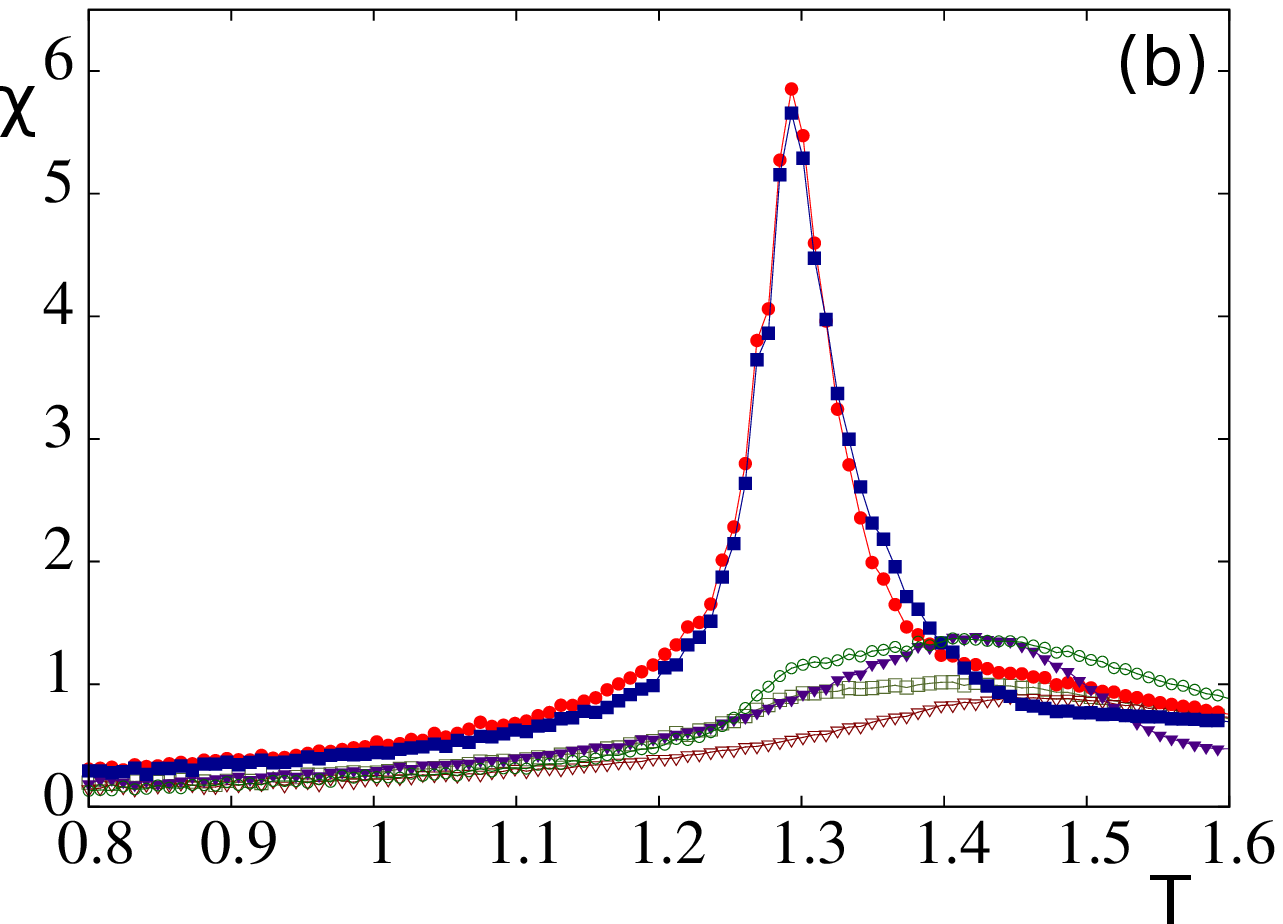}
\caption{(a) Layer magnetization and (b) layer magnetic susceptibility versus $T$
for  $H=0.2$, $J_2=-1$, $N_z=12$. Dark olive green void squares for the first layer, maroon void
triangles for the second, red circles for the third, indigo triangles for the fourth, dark blue squares for the fifth, dark green void circles for the sixth layer. \label{MXT2} }
\end{figure}

In an applied field, as seen earlier, in the GS all layers do not have the same characteristics so one expects different behaviors.  Figure \ref{MXT2} shows the layer magnetizations and the layer susceptibilities as functions of $T$ for  $H=0.2$ with $J_2=-1$, $N_z=12$ (only the first six layers are shown, the other six are symmetric).  Several remarks are listed below:

\begin{itemize}

\item Only layer 3 and layer 5  have a phase transition: their magnetizations strongly fall down  at the transition temperature. This can be understood from what we have anticipated above: these layers have the largest $xy$ components (see Fig. \ref{szsxy1}c).  Since the correlation between $xy$ components do not depend on the applied field, the temperature destroys the in-plane ferromagnetic ordering causing the transition. It is not the case for the $z$ components which are kept non zero by the field.  Of course, symmetric layers 8 and 10 have the same transition (not shown).

\item Layers with small amplitudes of $xy$ components do not have a strong transverse ordering at finite $T$: the absence of pronounced peaks in the susceptibility indicates that they do not make  a transition (see Fig. \ref{MXT2}).

\item Note that the $xy$ spin components of layers 3 and 5 are disordered at $T_c\simeq 1.275$ indicated by pronounced peaks of the susceptibility.
\end{itemize}

\begin{figure}[ht!]
\centering
\includegraphics[width=6cm,angle=0]{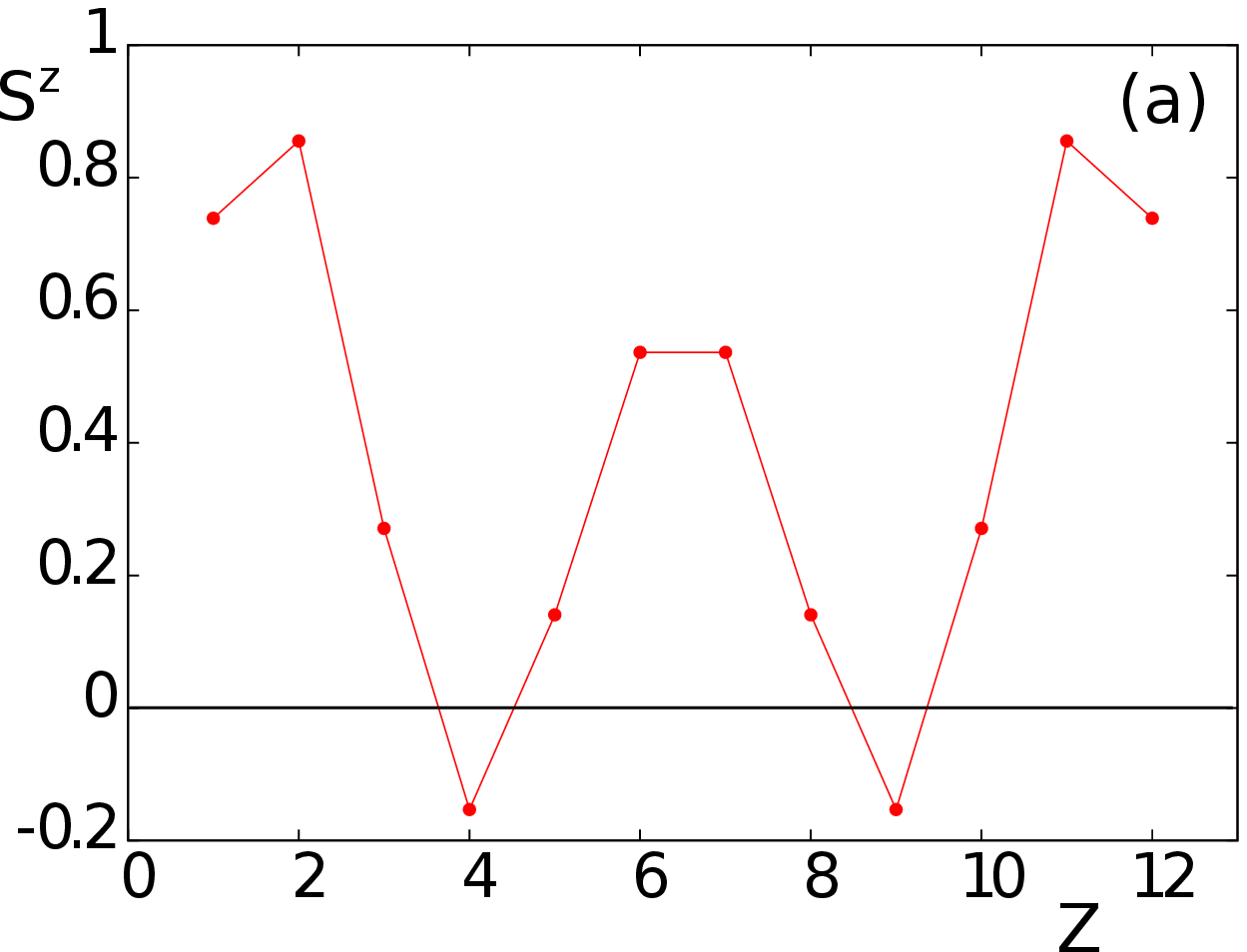}
\includegraphics[width=6cm,angle=0]{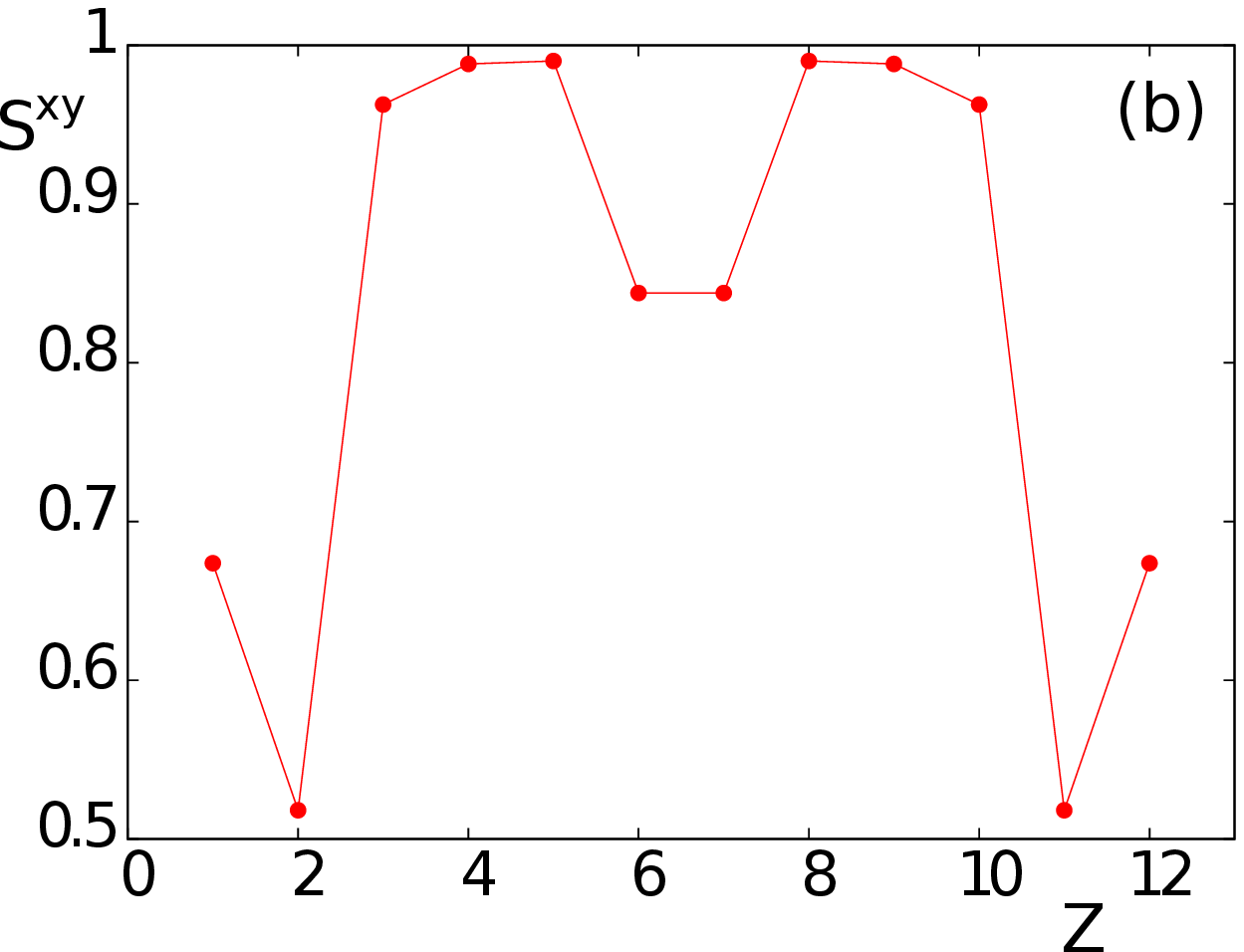}
\caption{(a) $S^{z}$ and (b) $S^{xy}$ across the film with  $H=0.7$. \label{szsxy2} }
\end{figure}

\begin{figure}[ht!]
\centering
\includegraphics[width=6cm,angle=0]{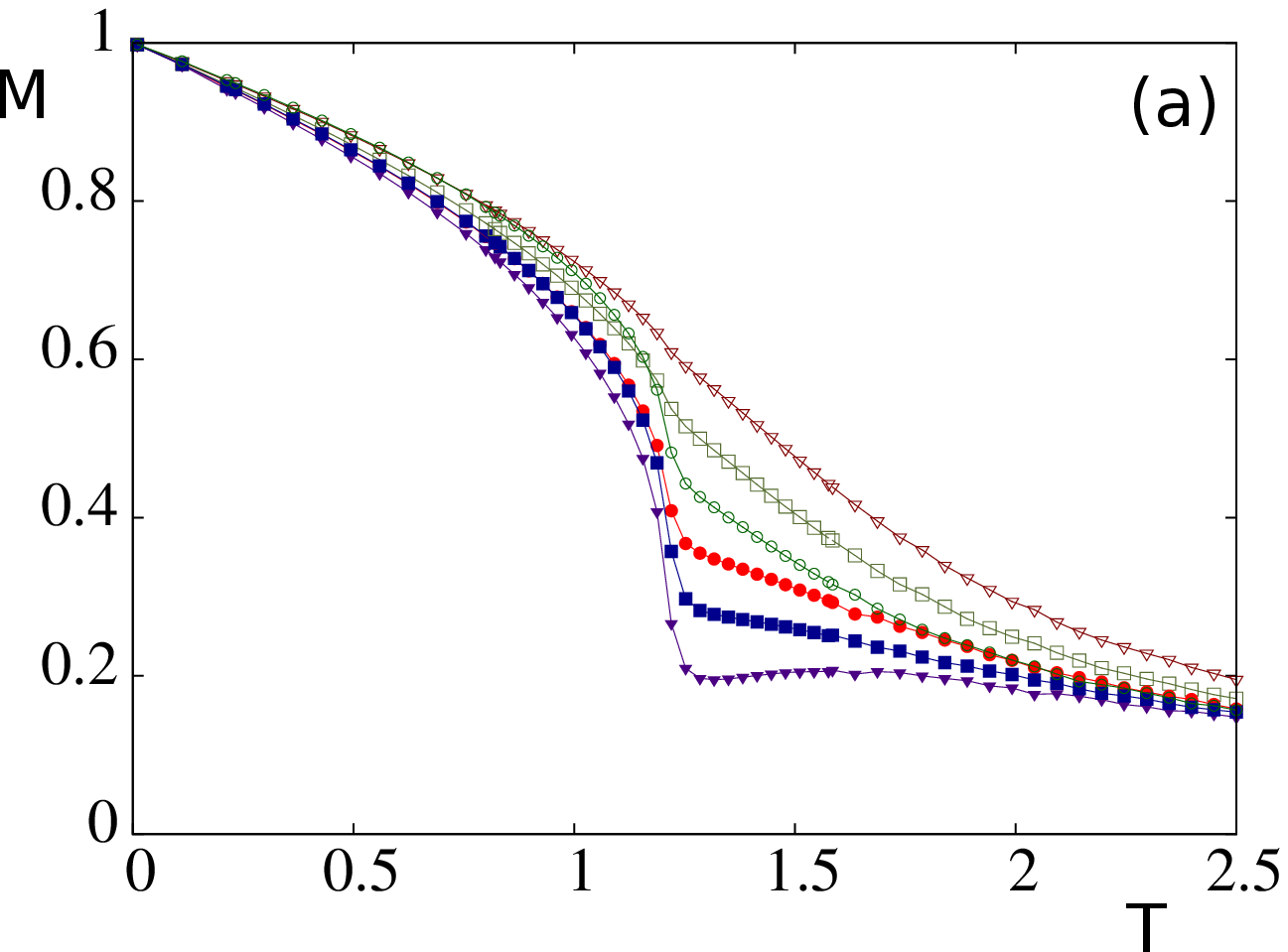}
\includegraphics[width=6cm,angle=0]{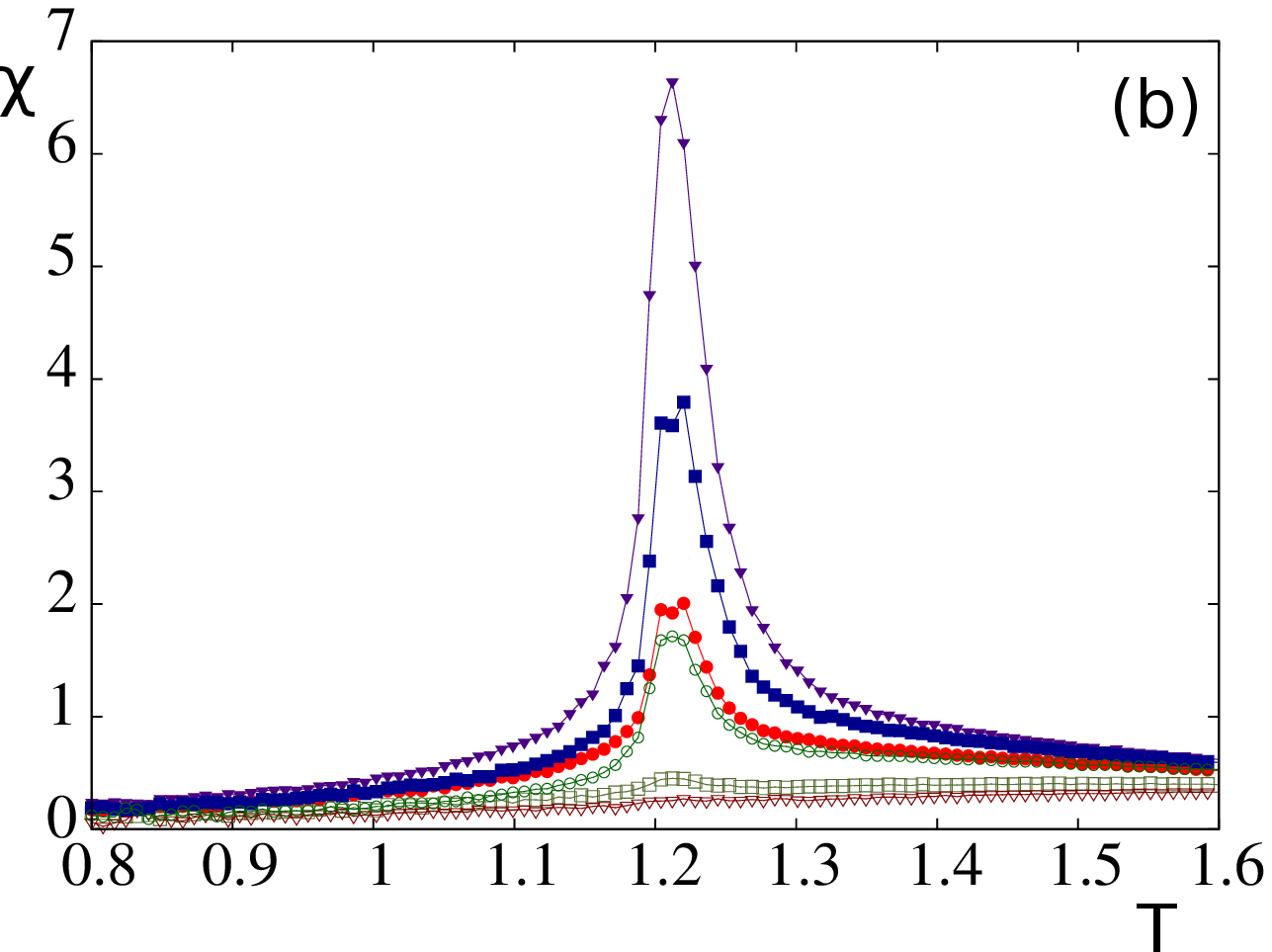}
\caption{(a) Layer magnetization and (b) layer magnetic susceptibility versus $T$
for  $H=0.7$, $J_2=-1$, $N_z=12$. Dark olive green void squares for the first layer, maroon void
triangles for the second, red circles for the third, indigo triangles for the fourth, dark blue squares for the fifth, dark green void circles for the sixth layer. \label{MXT4} }
\end{figure}

What we learn from the example shown above is that under an applied magnetic field the film can have a partial transition: some layers with large $xy$ spin components undergo a phase transition (destruction of their transverse $xy$ correlation). This picture is confirmed by several simulations for various field strengths. Another example is shown in the case of a strong field $H=0.7$: the GS is shown in Fig. \ref{szsxy2} where we observe large $xy$ spin components of layers 3, 4, and 5 (and symmetric layers 7, 8 and 9).
We should expect a transition for each of these layers. This is indeed the case: we show these transitions in Fig. \ref{MXT4} where sharp peaks of the susceptibilities of these layers are observed.
%Fig7

%Fig8

\begin{figure}[ht!]
\centering
\includegraphics[width=6cm,angle=0]{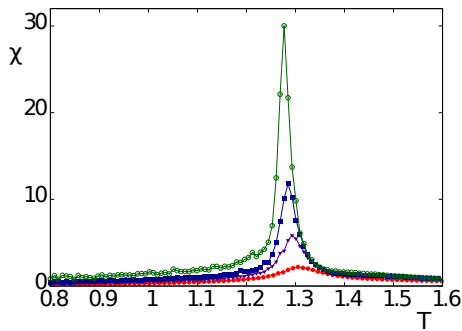}
\caption{Magnetic susceptibility of the third layer versus $T$ for $H=0.2$, $J_2=-1$, $N_z=12$. Dark green void circles, dark blue squares,
indigo triangles, red circles are susceptibilities for layer lattice sizes 100$\times$100, 60$\times$60, 40$\times$40 and 20$\times$20, respectively.\label{lateralsize} }
\end{figure}

We close this section by showing some size effects. Figure \ref{lateralsize} shows the effect of lateral size ($xy$ planes) on the layer susceptibility. As expected in a continuous transition, the peaks of the susceptibilities of the layers undergoing a transition grow strongly with the layer lattice size.

\subsection{Effects of the film thickness}

\begin{figure}[ht!]
\centering
\includegraphics[width=6cm,angle=0]{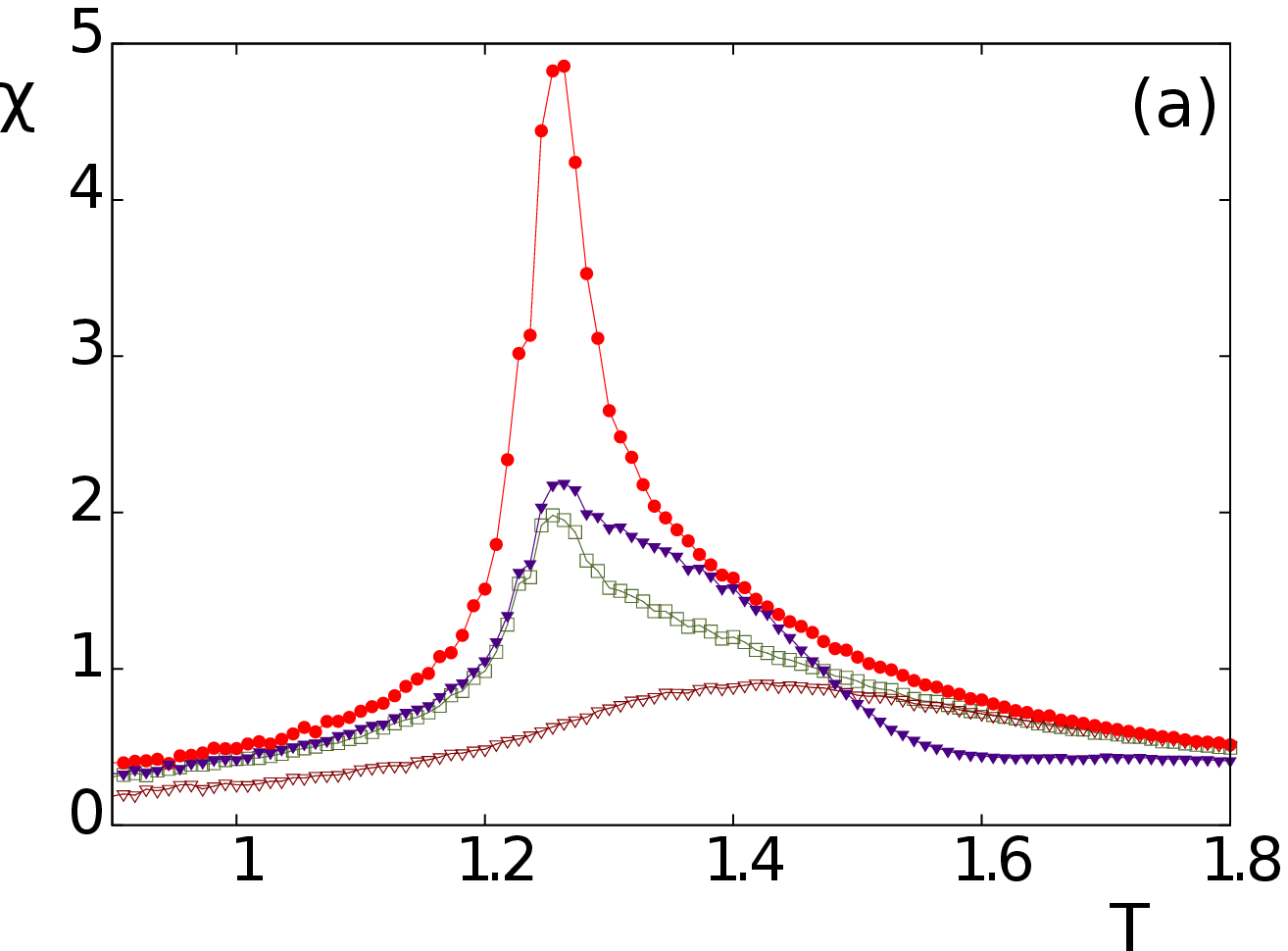}
\includegraphics[width=6cm,angle=0]{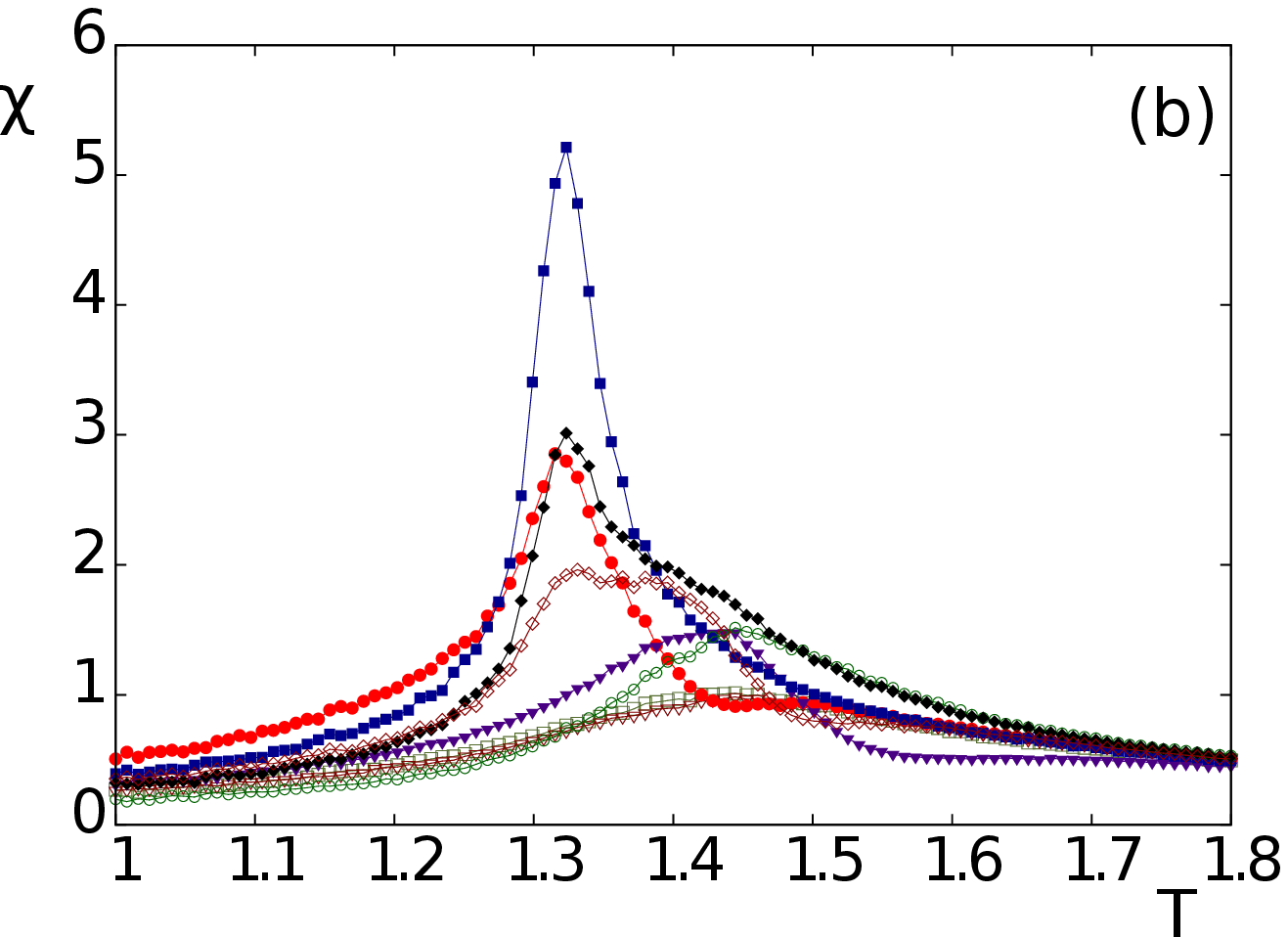}
\caption{Magnetic susceptibility versus $T$ for two thicknesses
with $H=0.2$, $J_2=-1$: (a) $N_z=8$, (b) $N_z=16$.
Dark olive green void squares are for the first layer, maroon void
triangles for the second, red circles for the third, indigo triangles for the fourth, dark blue squares for the fifth, dark green void circles for the sixth, black diamonds for the seventh, dark brown void diamonds for the eighth.  See text for comments.\label{Nzsize} }
\end{figure}

As for the thickness effects, we note that changing the thickness (odd or even number of layers) will change the GS spin configuration so that the layers with largest $xy$ components are not the same. As a consequence, the layers which undergo the transition are not the same for different thicknesses. We show in Fig. \ref{Nzsize} the layer susceptibilities for $N_z=8$ and 16.
For $N_z=8$, the layers which undergo a transition are the first, third and fourth layers with pronounced peaks, while for $N=16$, the layers which undergo a transition are the third, fifth, seventh and eighth layers.
%Fig9

%Fig10

\begin{figure}[ht!]
\centering
\includegraphics[width=6cm,angle=0]{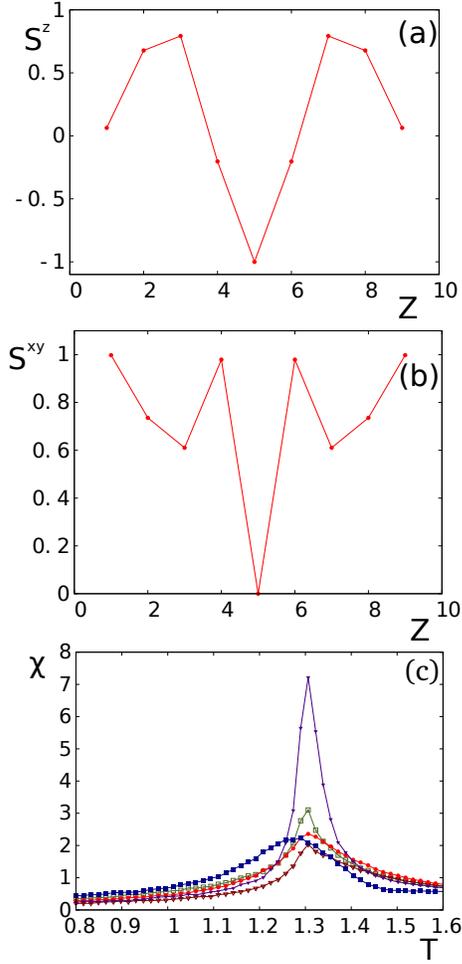}
\caption{ 9-layer film: spin components across the film in the case where $H=0.2$. The horizontal axis $Z$ represents plane $Z$ ($Z=1$ is the first plane etc.): (a) $S^{z}$ ; (b) Modulus $S^{xy}$ of the projection of the spins on the $xy$ plane ; (c) Layer susceptibilities versus $T$: Dark olive green void squares are for the first layer, maroon void
triangles for the second, red circles for the third, indigo triangles for the fourth, dark blue squares for the fifth layer, respectively. Only layers 4 and 1 (also layers 9 and 6, not shown) undergo a transition. See text for comments. \label{ffig11}}
\end{figure}

Let us show the case of an odd number of layers. Figure \ref{ffig11} shows the results for $N_z=9$ with $H=0.2$, $J_2=-1$. Due to the odd layer number, the center of symmetry is the middle layer (5th layer).  As seen, the layers 1 and 4 and their symmetric counterparts (layers 9 and 6) have largest $xy$ spin modulus (Fig. \ref{ffig11}b). The transition argument shown above predicts that these layers have a transversal phase transition in these $xy$ planes. This is indeed seen in Fig. \ref{ffig11}c where the susceptibility of layer 4 has a strong peak at the transition. The first layer, due to the lack of neighbors, has a weaker peak. The other layers do not undergo a transition. They show only a rounded maximum.

%Fig11

\section{Quantum fluctuations, layer magnetizations and spin-wave spectrum}\label{GFLT}
 We shall extend here the method used in Ref. \onlinecite{Diep2015} for zero field to the case where an applied magnetic field is present. The method remains essentially the same except the fact that each spin is defined not only by its angles with the NN in the adjacent layers but also by its azimuthal angle formed with the $c$ axis as seen in section \ref{GSSC}.

We use in the following the Hamiltonian (\ref{eqn:hamil1}) but with quantum Heisenberg spins $\mathbf S_i$ of magnitude $1/2$.  In addition,
it is known that in two dimensions there is no long-range order at finite temperature for isotropic spin models \cite{Mermin} with short-range interaction.  Since our films have small thickness, it is useful to add an anisotropic interaction to stabilize the long-range ordering at finite temperatures.
Let us use the following anisotropy between $\mathbf S_i$
and $\mathbf S_j$ which stabilizes the angle between their local quantization axes $S^z_i$ and $S^z_j$:
\begin{equation}
\mathcal H_a= -I_1\sum_{<i,j>}S^z_iS^z_j\cos\theta_{ij}
\end{equation}
where $I_1$ is supposed to be positive, small compared to $J_1$, and limited to NN.

The general method has been recently described in details in Refs. \onlinecite{Diep2015,Sahbi}. To save space, let us give the results for the simple cubic helimagnetic film in a field.
We define the following two double-time Green's functions in the real space:
\begin{eqnarray}
G_{i,j}(t,t')&=&<<S_i^+(t);S_{j}^-(t')>>\nonumber\\
&=&-i\theta (t-t')
<\left[S_i^+(t),S_{j}^-(t')\right]> \label{green59a}\\
F_{i,j}(t,t')&=&<<S_i^-(t);S_{j}^-(t')>>\nonumber\\
&=&-i\theta (t-t')
<\left[S_i^-(t),S_{j}^-(t')\right]>\label{green60}
\end{eqnarray}
Writing the equations of motion of these functions
and using the Tyablikov decoupling scheme to reduce the
higher-order functions, we
obtain the general equations for non collinear magnets \cite{Diep2015}. We next introduce the following in-plane Fourier
transforms $ g_{n,n'}$ and $  f_{n,n'}$ of the $G$ and $F$ Green's functions,
we finally obtain the following coupled equations
\begin{eqnarray}
&&D_n^-g_{n-2,n'}+E_n^-f_{n-2,n'}+B^-_{n}g_{n-1,n'}+C^-_{n}f_{n-1,n'}\nonumber\\
&&+(\omega+A_{n})g_{n,n'}+B^+_{n}g_{n+1,n'}
+ C^+_{n}f_{n+1,n'}+D_n^+g_{n+2,n'}\nonumber\\
&&+E_n^+f_{n+2,n'}=  2 \left< S^z_n\right>\delta_{n,n'}\label{eq:HGMatrixM1}\\
&&-E_n^-g_{n-2,n'}-D_n^-f_{n-2,n'}-C^-_{n}g_{n-1,n'}-B^-_{n}f_{n-1,n'}\nonumber\\
&&+(\omega-A_{n})f_{n,n'}-C^+_{n}g_{n+1,n'}-B^+_{n}f_{n+1,n'}\nonumber\\
&&-E_n^+g_{n+2,n'}-D_n^+f_{n+2,n'}=0\label{eq:HGMatrixM2}
\end{eqnarray}%
where $n=1,2,...,N_z$, $d_n=I_1/J_1^{\shortparallel}$, $\gamma=(\cos k_xa+\cos k_ya)/2$. The coefficients are given by
\begin{eqnarray}
A_{n} &=& - 8J_1^{\shortparallel}<S^z_n>(1+d_n-\gamma)\nonumber\\
&&-2 <S^z_{n+1}>\cos\theta_{n,n+1}(d_n+J_1^\bot)\nonumber\\
&&-2 <S^z_{n-1}>\cos\theta_{n,n-1}(d_n+J_1^\bot)\nonumber\\
&&-2J_2 < S^z_{n+2}>\cos\theta_{n,n+2}\nonumber\\
&&-2J_2 < S^z_{n-2}>\cos\theta_{n,n-2}-H\cos \zeta_n\label{Ancoeff}\nonumber\\
B_n^\pm &=& 2J_1^\bot \left< S^z_{n}\right>(\cos\theta_{n,n\pm 1}+1) \nonumber\\
C_n^\pm &=& 2J_1^\bot \left< S^z_{n}\right>(\cos\theta_{n,n\pm 1}-1)\nonumber\\
E_n^\pm &=& J_2 \left< S^z_{n}\right>(\cos\theta_{n,n\pm 2}-1)\nonumber\\
D_n^\pm &=& J_2 \left< S^z_{n}\right>(\cos\theta_{n,n\pm 2}+1) \nonumber
\end{eqnarray}
%MODIF END
$\omega$ is the spin-wave frequency, $k_{x}$ and  $k_{y}$
denote the wave-vector components in the $xy$ planes, $n$ is
the  index of the layer along the $c$ axis with $n=1$ being the surface layer,
$n=2$ the second layer and so on. The angle $\zeta_n$ is the azimuthal angle formed by a spin in the layer $n$ with the $c$ axis.
Note that (i) if $n=1$ then there are no  $n-1$ and $n-2$ terms in the matrix coefficients, (ii) if $n=2$ then there are no $n-2$ terms, (iii) if $n=N_z$ then there are no  $n+1$ and $n+2$ terms, (iv) if $n=N_z-1$ then there are no $n+2$ terms.  Besides, we have distinguished the in-plane NN interaction $J_1^{\shortparallel}$ from the inter-plane NN one $J_1^\bot$.
If we write all equations explicitly for $n=1,...,N_z$ we can put these equations under a matrix of dimension $2N_z\times 2N_z$. Solving this matrix equation, one gets the spin-wave frequencies $\omega$ at a given wave vector and a given $T$.

%\subsection{Results}
The layer magnetizations can be calculated at finite temperatures self-consistently.
The numerical method to carry out this task has been described in details in Ref. \onlinecite{Diep2015}.  It is noted that in bulk antiferromagnets and helimagnets the spin length is contracted at $T=0$ due to quantum fluctuations \cite{DiepTM}. Therefore, we also calculate the layer magnetization at $T=0$ \cite{Diep1979,Diep2015}.
It is interesting to note that due to the difference of the local field acting on a spin near the surface, the spin contraction is expected to be different for different layers.

\begin{figure}[ht!]
\centering
\includegraphics[width=6cm,angle=0]{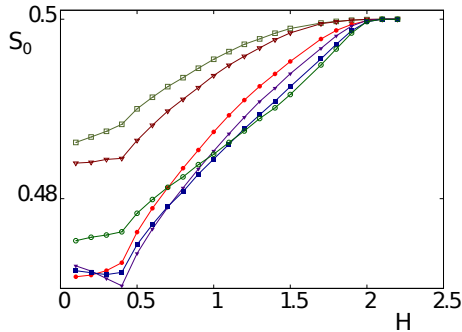}
\caption{Spin lengths at $T=0$ versus applied magnetic field $H$. Dark olive green void squares correspond to the spin length of the first layer, maroon void
triangles to that of the second, red circles to the third, indigo triangles to the fourth, dark blue squares to the fifth, dark green void circles to the sixth layer. \label{contract1}}
\end{figure}

\begin{figure}[ht!]
\centering
\includegraphics[width=6cm,angle=0]{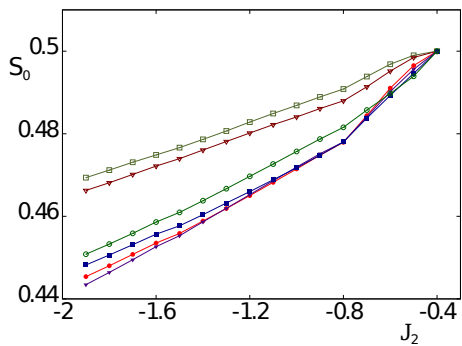}
\caption{Spin lengths at $T=0$ versus  $J_2$. Dark olive green void squares correspond to the spin length of the first layer, maroon void triangles to that of the second, red circles to the third, indigo triangles to the fourth, dark blue squares to the fifth, dark green void circles to the sixth layer. \label{contract2} }
\end{figure}

We show in Fig. \ref{contract1} the spin length of different layers at $T=0$ for $N=12$ and $J_2=-1$ as functions of $H$.
All spin contractions are not sensitive for $H$ lower than 0.4, but rapidly become smaller for further increasing $H$. They spin lengths are all saturated at the same value for $H>2$.
Figure \ref{contract2} shows the spin length as a function of $J_2$. When $J_2\geq -0.4$, the spin configuration becomes ferromagnetic, and as a consequence the contraction tends to 0. Note that in zero field, the critical value of $J_2$ is -0.25.
In both figures \ref{contract1} and \ref{contract2}, the surface layer and the third layer have smaller contractions than the other layers. This can be understood by examining the antiferromagnetic contribution to the GS energy of a spin in these layers: they are smaller than those of the other layers.
%Fig12
%Fig13

\begin{figure}[ht!]
\centering
\includegraphics[width=6cm,angle=0]{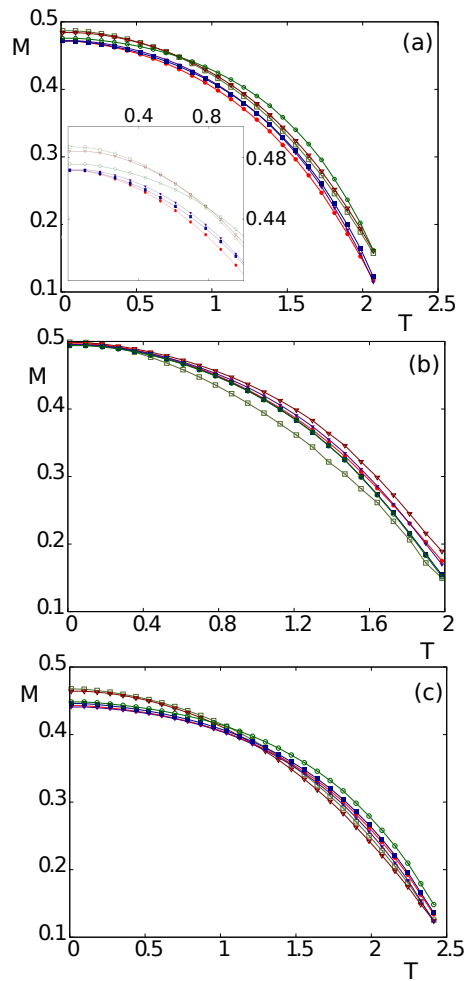}
\caption{Layer magnetizations versus $T$ for several values of $J_2$ with $H=0.2$, and $N_z=12$: (a)
$J_2=-1$, (b) $J_2=-0.5$, (c) $J_2=-2$. Dark olive green void squares correspond to the magnetization of the first layer, maroon void triangles to the second, red circles to the third, indigo triangles to the fourth, dark blue squares to the fifth, dark green void circles to the sixth layer. The inset in the top figure shows an enlarged region at low $T$. See text for comments.\label{lmT}}
\end{figure}

We show in Fig. \ref{lmT} the layer magnetizations versus $T$ for the case where $J_2=-1$ and $N_z=12$ (top figure).  The low-$T$ region is enlarged in the inset where one observes a crossover between the magnetizations of layers 1, 3 and 6 at $T\simeq 0.8$: below this temperature $M_1>M_3>M_6$ and above they become  $M_1<M_3<M_6$. This crossover is due to the competition between several complex factors: for example quantum fluctuations have less effect on the surface magnetization making it larger than magnetizations of interior planes at low $T$ as explained above (see Fig. \ref{contract1}), while the missing of neighbors for surface spins tends to diminish the surface magnetization at high $T$ \cite{DiepTF91,Diep2015}.
The middle figure shows the case where $J_2=-0.5$  closer to the ferromagnetic limit. The spin length at $T=0$ is almost 0.5 (very small contraction) and there is no visible crossover observed in the top figure. The bottom figure shows the case $J_2=-2$ which is the case of a strong helical angle. We observe then a crossover  at a higher $T$ ($\simeq 1.2$) which
is in  agreement with the physical picture given above on the competition between quantum and thermal fluctuations.
Note that we did not attempt to get closer to the transition temperature, namely $M<0.1$  because the convergence of the self-consistency then becomes bad.
%Fig14

\begin{figure}[ht!]
\centering
\includegraphics[width=6cm,angle=0]{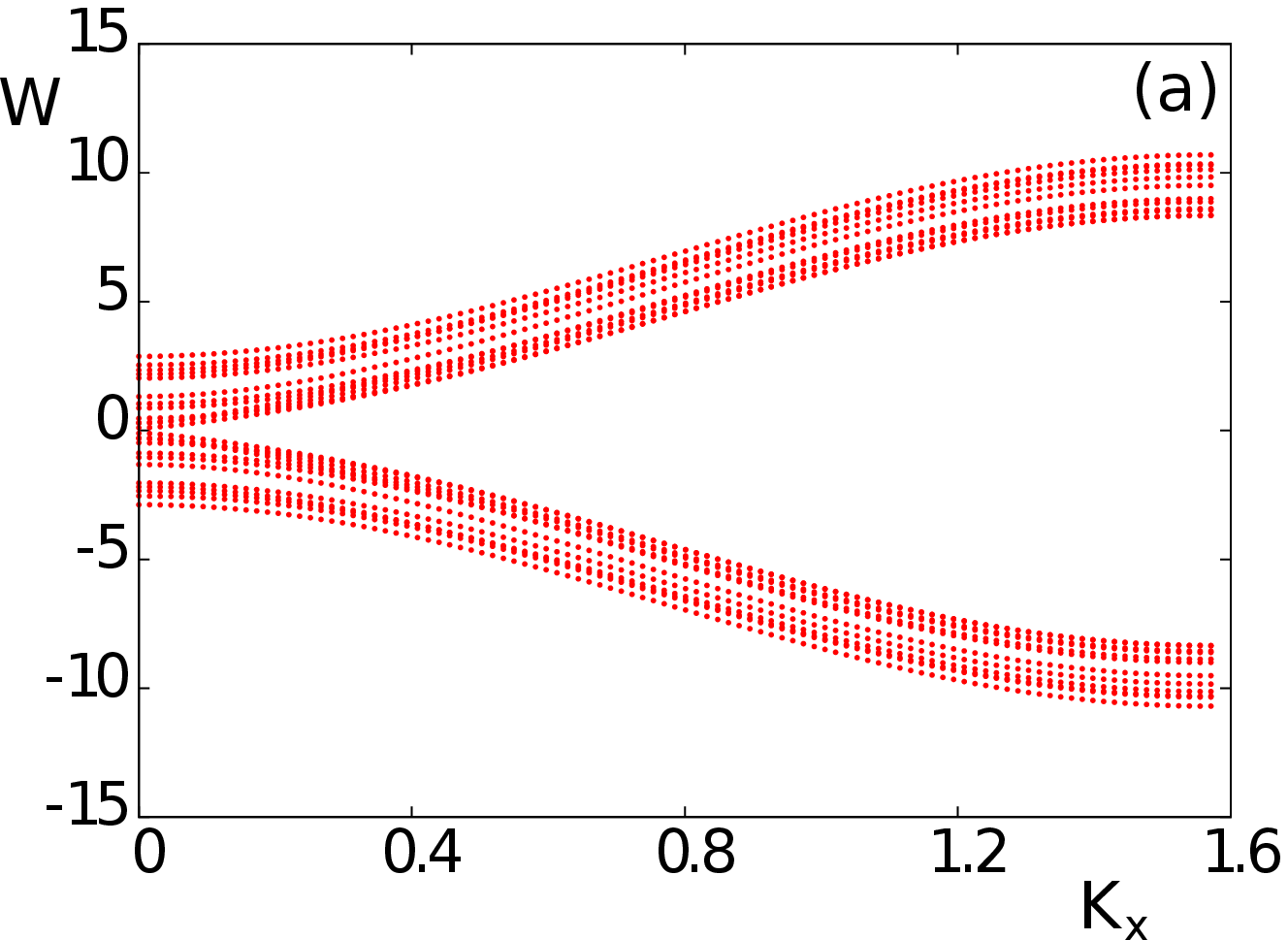}
\includegraphics[width=6cm,angle=0]{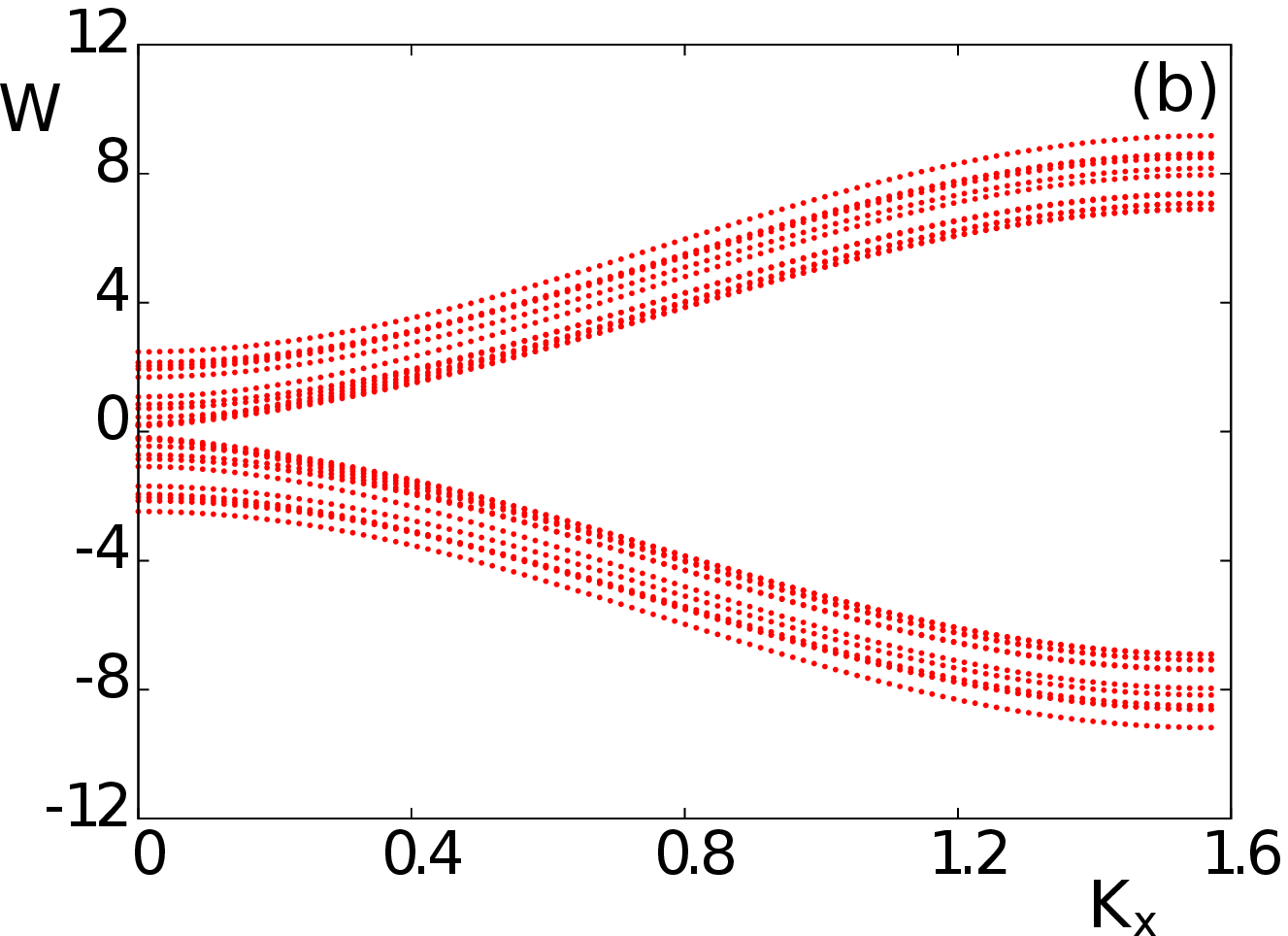}
\caption{Spin-wave spectrum  versus $k_x=k_y$ where $W$ stands for spin-wave frequency $\omega$ in Eqs. (\ref{eq:HGMatrixM1})-(\ref{eq:HGMatrixM2}), at (a) $T=0.353$ and (b) $T=1.212$, with $H=0.2$.\label{sw1}}
\end{figure}

\begin{figure}[ht!]
\centering
\includegraphics[width=6cm,angle=0]{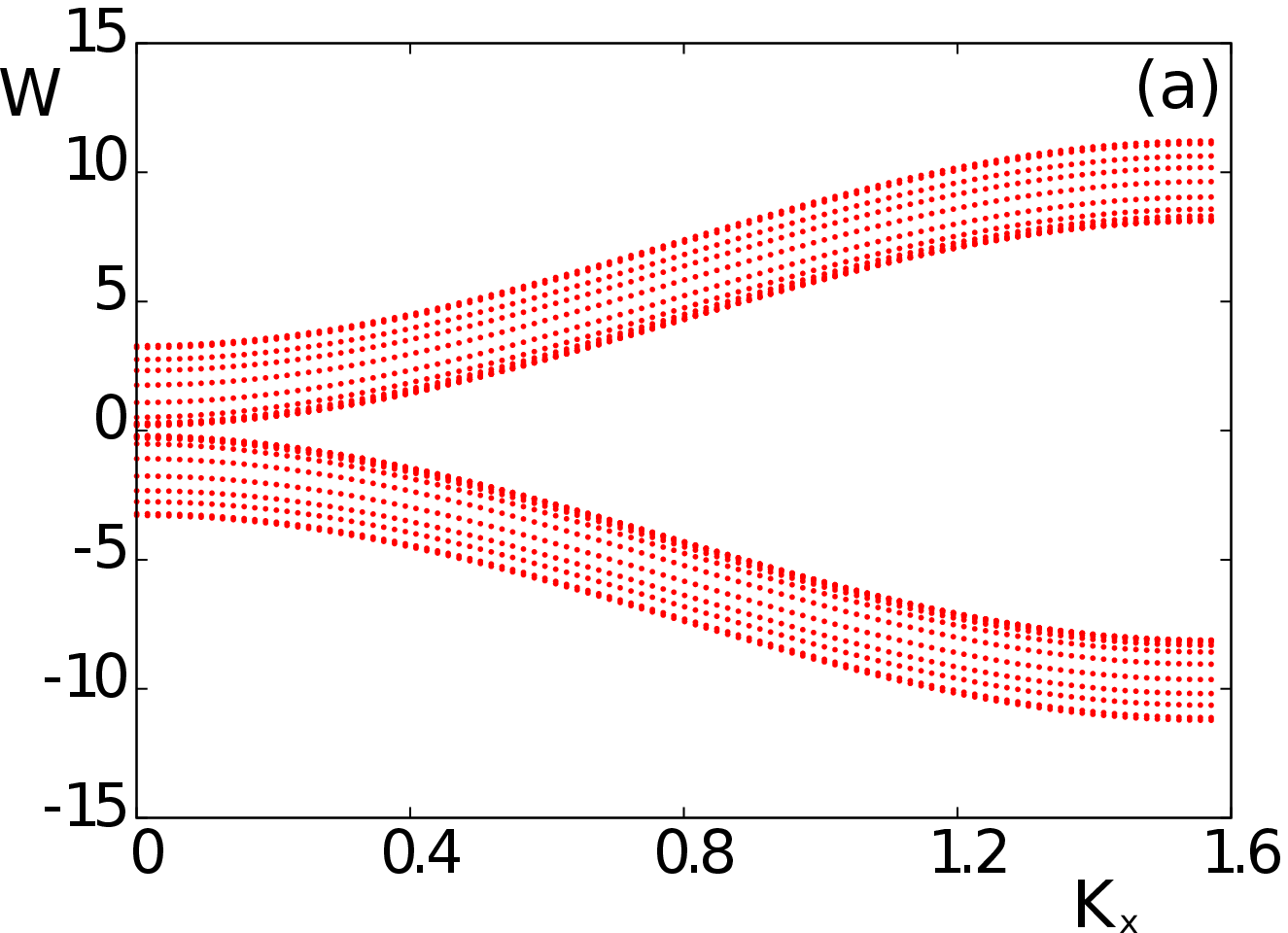}
\includegraphics[width=6cm,angle=0]{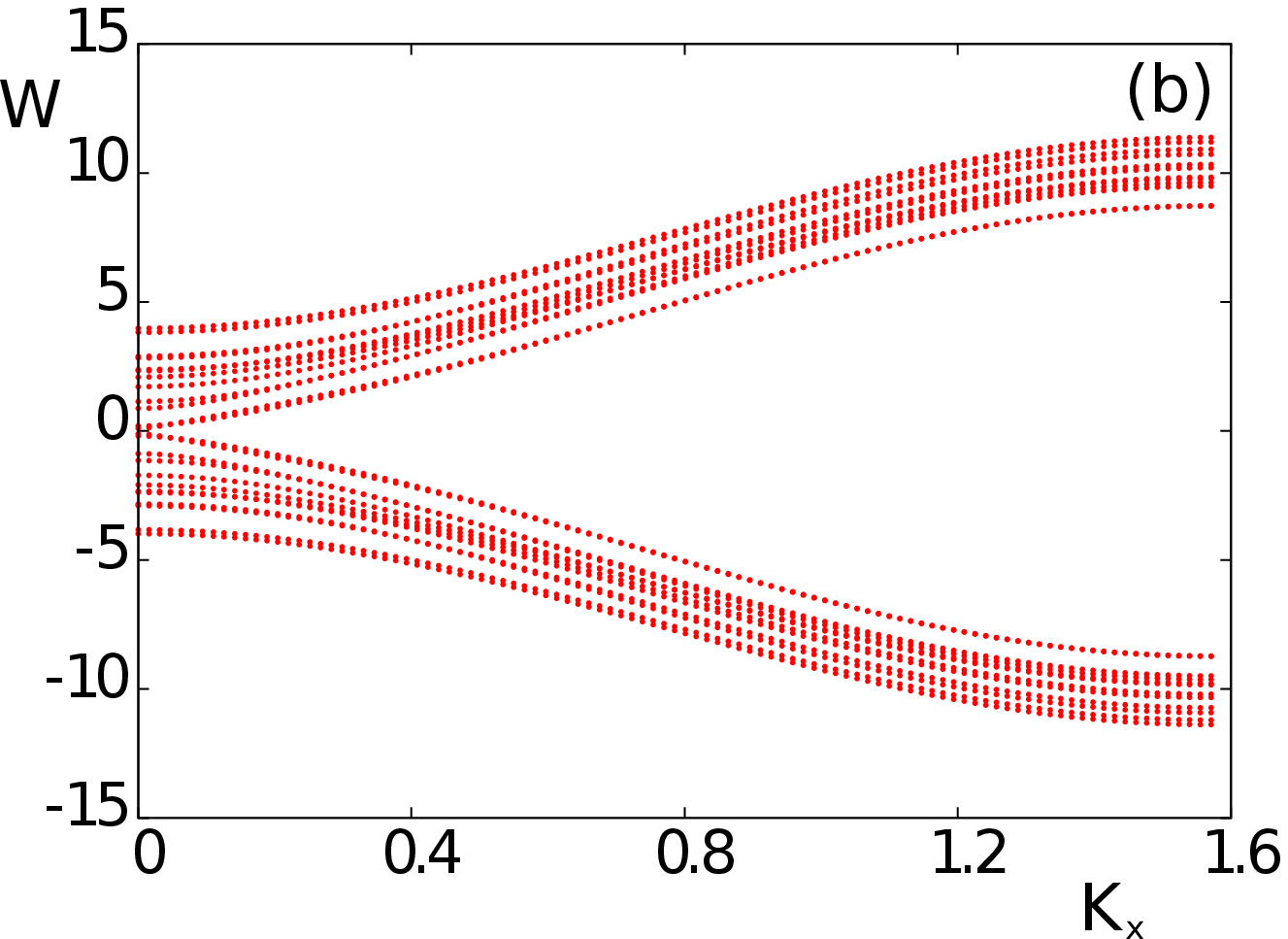}
\caption{Spin-wave spectrum versus $k_x=k_y$ at $T=0.353$ where $W$ stands for spin-wave frequency $\omega$ in Eqs. (\ref{eq:HGMatrixM1})-(\ref{eq:HGMatrixM2}),  for (a) $J_2=-0.5$ and (b) $J_2=-2$, with $H=0.2$. \label{sw2}}
\end{figure}

Before closing this section, let us discuss about the spin-wave spectrum.  Let us remind that to solve self-consistently Eqs. (\ref{eq:HGMatrixM1})-(\ref{eq:HGMatrixM2}) at each $T$, we use as inputs $<S_1^z>,<S_2^z>, ..., <S_{N_z}^z>$ to search for the eigenvalues $\omega$ for each vector $(k_x,k_y)$ and then calculate the outputs  $<S_1^z>,<S_2^z>, ..., <S_{N_z}^z>$. The self-consistent solution is obtained when the outputs are equal to the inputs at a desired convergence precision fixed at the fifth digit, namely  $10^{-5}$ (see other details in Ref. \onlinecite{Diep2015}).  Figure \ref{sw1} shows the spin-wave spectrum  in the direction $k_x=k_y$ of the Brillouin zone at $T=0.353$ and $T=1.212$ for comparison.  As seen, as $T$ increases the spin-wave frequency decreases. Near the transition (not shown) it tends to zero.  Figure \ref{sw2} shows the spin-wave spectrum at $T=0.353$ for $J_2=0.5$ and $J_2=-1$, for comparison. Examining them closely, we see that the distribution of the spin-wave modes (positions of the branches in the spectrum) are quite different for the two cases. When summed up for calculating the layer magnetizations, they give rise to the difference observed for the two cases shown in Fig. \ref{lmT}.

%Fig15

%Fig16

\section{Concluding remarks}
In this paper we have shown (i) the GS spin configuration of a Heisenberg helimagnetic thin film in a magnetic field applied along the $c$ axis perpendicular to the film, (ii) the phase transition occurring in the film at a finite temperature, (iii) quantum effects at low $T$ and the temperature dependence of the layer magnetizations as well as the spin-wave spectrum.

Synthetically, we can summarize that under the applied magnetic field,  the spins in the GS make different angles between them across the film. When the temperature increases, the layers with large $xy$ spin-components undergo a  phase transition where the transverse (in-plane) $xy$ ordering is destroyed.  This "transverse" transition is possible because the $xy$ spin-components do not depend on the field.  Other layers with small $xy$ spin-components, namely large $z$ components, do not make a transition because the ordering in $S^z$ is maintained by the applied field.  The transition of a number of layers with large $xy$ spin-components, not all layers, is a new phenomenon discovered here with our present model.

We have also investigated the quantum version of the model by using the Green's function method. The results show that the zero-point spin contraction is different from layer to layer. We also find a crossover of layer magnetizations which depends on $J_2$, namely on the magnitude of helical angles.

Experiments are often performed on materials with  helical structures often more complicated than the model considered in this paper. However, the clear physical pictures given in our present analysis are believed to be useful in the search for the interpretation of experimental data.

\acknowledgments
SEH acknowledges  a financial support from Agence Universitaire de la Francophonie (AUF).

{}


\begin{thebibliography}{0}
\expandafter\ifx\csname natexlab\endcsname\relax\def\natexlab#1{#1}\fi
\expandafter\ifx\csname bibnamefont\endcsname\relax
  \def\bibnamefont#1{#1}\fi
\expandafter\ifx\csname bibfnamefont\endcsname\relax
  \def\bibfnamefont#1{#1}\fi
\expandafter\ifx\csname citenamefont\endcsname\relax
  \def\citenamefont#1{#1}\fi
\expandafter\ifx\csname url\endcsname\relax
  \def\url#1{\texttt{#1}}\fi
\expandafter\ifx\csname urlprefix\endcsname\relax\def\urlprefix{URL }\fi
\providecommand{\bibinfo}[2]{#2}
\providecommand{\eprint}[2][]{\url{#2}}

\end{thebibliography}


\begin{thebibliography}{9}


\bibitem{Yoshimori} A. Yoshimori, J. Phys. Soc. Jpn {\bf 14}, 807 (1959).
\bibitem{Villain59} J. Villain, Phys. Chem. Solids {\bf 11}, 303 (1959).
\bibitem{Bertaut1} E.  F.  Bertaut,  Configurations  de  Spin  et  Th\'eorie  des Groups, Journal de Physique et le Radium {\bf 22}, 321 (1961).

\bibitem{Bertaut2} E.  F.  Bertaut,  Lattice  Theory  of  Spin  Configuration, J. Applied  Phys. {\bf  33}, 1138 (1962).

\bibitem{Bertaut3} A. Kallel, H. Boller and E. F. Bertaut, Helimagnetism in
MnP-Type Compounds: MnP, FeP, CrAs and CrAs(1-x)Sb(x) Mixed  Crystals,
J. Phys. and  Chem. Solids {\bf 35}, 1139 (1974).

\bibitem{Harada} I. Harada and K. Motizuki, J. Phys. Soc. Jpn {\bf 32}, 927 (1972).
\bibitem{Rastelli} E. Rastelli, L. Reatto and A. Tassi, Quantum fluctuations in helimagnets, J. Phys. C {\bf 18}, 353 (1985).
\bibitem{Diep89} H. T. Diep, Low-temperature properties of quantum Heisenberg helimagnets, Phys. Rev. B {\bf 40}, 741 (1989).
\bibitem{Quartu1998} R. Quartu and H. T. Diep, Phase diagram of  body-centered tetragonal Helimagnets, J. Magn. Magn. Mater. {\bf 182}, 38 (1998).
\bibitem{Stishov}S. M. Stishov, A. E. Petrova, S. Khasanov, G. Kh. Panova, A. A. Shikov, J. C. Lashley, D. Wu, and T. A. Lograsso,
Magnetic phase transition in the itinerant helimagnet MnSi: Thermodynamic and transport properties,
Phys. Rev. B {\bf 76}, 052405 (2007).


\bibitem{Bak} P. Bak and M. H. Jensen, J. Phys. C {\bf 13}, L881 (1980).

\bibitem{Plumer} M. L. Plumer and M. B. Walker, J. Phys. C {\bf 14}, 4689 (1981).

\bibitem{Maleyev} Maleyev, Phys. Rev. B {\bf 73}, 174402 (2006).

%\bibitem{Fert2013} A. Fert, V. Cros, and J. Sampaio,
%Nat. Nanotechnol. {\bf 8}, 152 (2013).
%\bibitem{Lin} Shi-Zeng Lin, Avadh Saxena, and Cristian D. Batista, Phys. Rev. B {\bf 91}, 224407 (2015).
%\bibitem{Bogdanov} A. N. Bogdanov and D. A. Yablonskii, Sov. Phys. JETP
%{\bf 68}, 101 (1989).
%\bibitem{Rossler} U.K.R\"{o}ßler, A. N. Bogdanov, and C. Pfleiderer,
%Nature (London) {\bf 442}, 797 (2006).
%\bibitem{Muhlbauer} S. M\"{u}hlbauer, B. Binz, F. Jonietz, C. Pfleiderer, A. Rosch, A. Neubauer, R. Georgii, and P. B\"{o}ni, Science {\bf 323}, 915 (2009).
%\bibitem{Yu1} X. Z. Yu, Y. Onose, N. Kanazawa, J. H. Park, J. H. Han, Y.
%Matsui, N. Nagaosa, and Y. Tokura, Nature (London) {\bf 465}, 901 (2010).
%\bibitem{Yu2} X. Z. Yu, N. Kanazawa, Y. Onose, K. Kimoto, W. Z. Zhang,
%S. Ishiwata, Y. Matsui, and Y. Tokura, Nat. Mater. {\bf 10}, 106 (2011).
%\bibitem{Seki} S. Seki, X. Z. Yu, S. Ishiwata, and Y. Tokura,
%Science {\bf 336}, 198 (2012).
%\bibitem{Adams} T. Adams, A. Chacon, M. Wagner, A. Bauer, G. Brandl, B.
%Pedersen, H. Berger, P. Lemmens, and C. Pfleiderer,
%Phys. Rev. Lett. {\bf 108}, 237204 (2012).
%
%\bibitem{Heurich} J. Heurich, J. K\"{o}nig, and A. H. MacDonald, Phys. Rev. B {\bf 68}, 064406 (2003).
%
%\bibitem{Wessely} O. Wessely, B. Skubic, and L. Nordstrom, Phys. Rev. B {\bf 79}, 104433 (2009).
%
%\bibitem{Jonietz} F. Jonietz, S. M\"{u}hlbauer, C. Pfleiderer, A. Neubauer, W. Munzer, A. Bauer, T. Adams, R. Georgii, P. B\"{o}ni, R. A. Duine, K. Everschor, M. Garst, and A. Risch, Science {\bf 330}, 1648 (2010).
%
%
%\bibitem{Hayami2016} S. Hayami, S-Z. Lin, C. D. Batista, Bubble and skyrmion crystals in frustrated magnets with easy-axis anisotropy, Phys. Rev. B {\bf 93}, 184413 (2016).
%
%\bibitem{Okubo} T. Okubo, S. Chung and H. Kawamura, Phys. Rev. Lett.  {\bf 108}, 017206 (2012).

\bibitem{DiepFSS} H. T. Diep (ed.), {\it Frustrated Spin Systems}, 2nd edition, World Scientific (2013).

\bibitem{NgoSurface} V. Thanh Ngo and H. T. Diep, Effects of frustrated surface in Heisenberg thin films,
Phys. Rev. B {\bf 75}, 035412 (2007), Selected for the Vir. J. Nan. Sci. Tech. {\bf 15}, 126 (2007).

\bibitem{NgoSurface2} V. Thanh Ngo and H. T. Diep, Frustration effects in antiferrormagnetic face-centered cubic Heisenberg films,
J. Phys: Condens. Matter. {\bf 19}, 386202 (2007).


\bibitem{Diep2015} H. T. Diep, Quantum Theory of Helimagnetic Thin Films, Phys. Rev. B {\bf 91}, 014436 (2015).


\bibitem{Sahbi} Sahbi El Hog and H. T. Diep, Helimagnetic Thin Films: Surface Reconstruction, Surface Spin-Waves, Magnetization, J. Magn. and Magn. Mater.  400, 276-281 (2016).
\bibitem{DiepTM} See for example H. T. Diep, {\it Theory of Magnetism: Application to Surface Physics}, World Scientific, Singapore (2014).


\bibitem{Diep89b} H. T. Diep, Magnetic transitions in helimagnets,
Phys. Rev. B {\bf 39}, 397 (1989).

\bibitem{Ngo08}   V. Thanh Ngo and H. T.  Diep, Stacked triangular XY antiferromagnets: End of a controversial issue on the phase transition, J. Appl. Phys. {\bf 103}, 07C712 (2007).

\bibitem{Ngo09} V. Thanh Ngo and H. T. Diep, Phase transition in Heisenberg stacked triangular antiferromagnets: End of a controversy, Phys. Rev. E {\bf 78}, 031119 (2008).




\bibitem{Heinrich} {\it Ultrathin Magnetic Structures}, vol. I and II, J.A.C. Bland and B. Heinrich (editors), Springer-Verlag (1994).
\bibitem{Zangwill} A. Zangwill, {\it Physics at Surfaces}, Cambridge University Press (1988).

\bibitem{Mello2003} V. D. Mello, C. V. Chianca, Ana L. Danta, and A. S. Carri\c{c}, Magnetic surface phase of thin helimagnetic films, Phys. Rev. B {\bf 67}, 012401  (2003).
\bibitem{Cinti2008} F. Cinti, A. Cuccoli, and A. Rettori, Exotic magnetic structures in ultrathin helimagnetic holmium films, Phys. Rev. B {\bf 78}, 020402(R) (2008).
\bibitem{Karhu2011} E. A. Karhu, S. Kahwaji, M. D. Robertson, H. Fritzsche, B. J. Kirby, C. F. Majkrzak, and T. L. Monchesky, Helical magnetic order in MnSi thin films, Phys. Rev. B {\bf 84}, 060404(R) (2011).
\bibitem{Karhu2012} E. A. Karhu, U. K. R\"{o}$\beta$ler, A. N. Bogdanov, S. Kahwaji,  B. J. Kirby,  H. Fritzsche, M. D. Robertson, C. F. Majkrzak, and T. L. Monchesky, Chiral modulation and reorientation effects in MnSi thin films, Phys. Rev. B {\bf 85}, 094429 (2012).
\bibitem{Heurich} J. Heurich, J. K\"{o}nig, and A. H. MacDonald, Phys. Rev. B {\bf 68}, 064406 (2003).

\bibitem{Wessely} O. Wessely, B. Skubic, and L. Nordstrom, Phys. Rev. B {\bf 79}, 104433 (2009).

\bibitem{Jonietz} F. Jonietz, S. M\"{u}hlbauer, C. Pfleiderer, A. Neubauer, W. Munzer, A. Bauer, T. Adams, R. Georgii, P. B\"{o}ni, R. A. Duine, K. Everschor, M. Garst, and A. Risch, Science {\bf 330}, 1648 (2010).


\bibitem{Mermin} N. D. Mermin and H. Wagner, Phys. Rev. Lett. {\bf 17}, 1133 (1966).

\bibitem{Diep1979} Diep-The-Hung, J. C. S. Levy and O. Nagai, Effect of surface spin-waves and surface anisotropy in magnetic thin films at finite temperatures, Phys. Stat. Solidi (b) {\bf 93}, 351 (1979).
\bibitem{DiepTF91} H. T. Diep, Quantum effects in antiferromagnetic thin films,
Phys. Rev. B {\bf 43}, 8509 (1991).

%\bibitem{DiepSL89} H. T. Diep, Theory of antiferromagnetic superlattices at finite temperatures, Phys. Rev. B {\bf 40}, 4818 (1989).




%\bibitem{Leiner} V. Leiner, D. Labergerie, R. Siebrecht, Ch. Sutter, and H. Zabel, Physica B {\bf 283}, 167 (2000).





\end{thebibliography}
\end{document}